\documentclass[11pt,letterpaper]{article}
\usepackage{makeidx}
\usepackage{eurosym}
\usepackage{amsthm}
\usepackage{epsfig,lscape,rotate}
\usepackage{subcaption}
\usepackage{graphicx}
\usepackage{amsmath}
\usepackage{epstopdf}
\usepackage{lscape}
\usepackage{rotating}
\usepackage{amssymb}
\usepackage{amsfonts}
\usepackage{euscript}
\usepackage{mathrsfs}
\usepackage{natbib}
\usepackage{multirow}
\usepackage{geometry}
\usepackage{algorithm}
\usepackage{booktabs}
\usepackage[usenames]{color}
\usepackage{xr}
\usepackage{tabularx}
\usepackage{threeparttable}
\usepackage{adjustbox}
\usepackage{tikz}
\usepackage{float}
\usepackage[colorlinks=false]{hyperref}
\usepackage{lscape} 
\usepackage{url}
\usepackage{bm}
\usepackage{tabularx}
\usepackage{makecell}
\usepackage{bbding}
\usepackage{lscape}

\newtheorem{theorem}{Theorem}
\newtheorem{definition}{Definition}

\newtheorem{assumption}{Assumption}

\newtheorem{lemma}{Lemma}

\newtheoremstyle{ecta}
{\medskipamount}{\bigskipamount}{\normalfont}{1.4em}{\scshape}{:}{1em}{}
\theoremstyle{ecta}
\newtheorem*{example*}{Example}

\hypersetup{colorlinks=true, citecolor=blue, linkcolor=black}

\newcolumntype{Y}{>{\centering\arraybackslash}X}

\newcolumntype{b}{X}
\newcolumntype{s}{>{\hsize=.5\hsize}Y}
\def\sym#1{\ifmmode^{#1}\else\(^{#1}\)\fi}

\input{tcilatex}
\begin{document}

	\begin{titlepage}
		\title{Testing for the Minimum Mean-Variance Spanning Set\footnote{We thank Andrew Atkeson and Whitney Newey for their comments.}}
		
		\date{\today}
		\vspace{-1.0cm}
		\author{\mbox{}\\\mbox{}\\Zhipeng Liao\thanks{Department of Economics, UCLA, Los Angeles, CA 90095; e-mail: zhipeng.liao@econ.ucla.edu.}~~~~~~ Bin Wang\thanks{School of Economics and Management, Harbin Institute of Technology, Shenzhen, China; e-mail: wangbin81@hit.edu.cn.}~~~~~Wenyu Zhou\thanks{International Business School, Zhejiang  University, Haining, Zhejiang 314400, China; e-mail: wenyuzhou@intl.zju.edu.cn.}
			\\
			\mbox{}\\\mbox{} }
	\end{titlepage}
	
	\maketitle
	
	\begin{abstract}
		\thispagestyle{empty}\noindent This paper explores the estimation and inference of the minimum spanning set (MSS), the smallest subset of risky assets that spans the mean-variance efficient frontier of the full asset set. We establish identification conditions for the MSS and develop a novel procedure for its estimation and inference. Our theoretical analysis shows that the proposed MSS estimator covers the true MSS with probability approaching 1 and converges asymptotically to the true MSS at any desired confidence level, such as 0.95 or 0.99. Monte Carlo simulations confirm the strong finite-sample performance of the MSS estimator. We apply our method to evaluate the relative importance of individual stock momentum and factor momentum strategies, along with a set of well-established stock return factors. The empirical results highlight factor momentum, along with several stock momentum and return factors, as key drivers of mean-variance efficiency. Furthermore, our analysis uncovers the sources of contribution from these factors and provides a ranking of their relative importance, offering new insights into their roles in mean-variance analysis. 
		\newline
		\newline
		\noindent \textbf{Keywords}: Mean-variance efficiency, Moving block bootstrap, Set inference,  Spanning test. \newline
		\noindent \textbf{JEL Codes}: C14, C22.\newline
	\end{abstract}

\renewcommand{\arraystretch}{1.5}\setlength{\baselineskip}{6.5mm}\

\setcounter{equation}{0} \renewcommand{\theequation}{\thesection.%
\arabic{equation}}

\setcounter{page}{1}

\vspace{-1.0cm}


\section{Introduction\label{sec-intro}}

Conventional spanning tests, which assess whether one set of risky assets
can span another, have been proposed and widely utilized in asset management
and empirical research (e.g., \cite{huberman1987}, \cite%
{ferson-foerster-keim-93}, \cite{deroon-nijman-werker-01}, \cite%
{amengual2010}, \cite{kan2012tests}, \cite{penaranda2012}, among others). In
practice, these tests are often used to determine whether an additional set
of risky assets can further extend the mean-variance efficient frontier of a
given set of benchmark assets. Despite their extensive use, a notable
limitation remains: to the best of our knowledge, no existing methods can
estimate the smallest subset of assets that preserves the efficient frontier
of the full set, including both the benchmark and the additional assets.
This gap is significant given the growing demand among practitioners to
identify the most relevant assets.

To address this gap, we propose an estimation procedure for identifying the 
\emph{minimum spanning set} (\textrm{MSS}) within a given collection of
risky assets. Formally, consider a set of $d$ assets ($d\geq 2$)\
represented by their returns, $R=(R_{i})_{i\leq d}$. Our objective is to
assess whether the size of $R$ can be reduced without compromising its
mean-variance efficiency and to identify the smallest subset of assets that
reproduces the efficient frontier of the full set\ $R$. This subset,
referred to as the \textrm{MSS}, is the focus of this paper.

Our research question is related to, but distinct from, those addressed by
traditional spanning tests. Conventional tests evaluate whether an
additional set of assets, taken as a whole, is redundant, i.e., whether
adding these assets to a benchmark set extends its mean-variance efficient
frontier. However, they provide no insights into the relative importance of
individual assets within either the additional or benchmark set, nor do they
address whether any subsets within these groups are redundant and can be
excluded without compromising mean-variance efficiency.

In contrast, our method directly estimates the \textrm{MSS} and offers
statistical insights into the relative importance of assets within the
entire set. Additionally, when new assets beyond $R$ become available, our
method evaluates their relevance and determines whether their inclusion
renders any existing assets in $R$ redundant. This approach provides valid
statistical inference on asset relevance, and is valuable for investors who
aim to minimize asset management costs by identifying and investing in the
smallest subset of assets capable of maintaining mean-variance efficiency.

To ensure that the estimation and inference of the \textrm{MSS} is
well-defined, we begin by establishing its existence and uniqueness under
the mild assumption that the variance-covariance matrix of $R$ is
non-singular.\ Next, we derive the identification conditions for the \textrm{%
MSS}\ based on a set of restrictions on the regression coefficients. These
coefficients depend exclusively on the first two moments of $R$, ensuring
they are consistently estimable. Consequently, the restrictions embedded in
the identification conditions are empirically testable.

We construct a statistic $M_{i,T}$, where $T$ denotes the sample size, to
evaluate the identification restrictions for each asset $R_{i}$ in $R$. This
statistic converges in distribution to a maximum normal distribution if $%
R_{i}$ is redundant and diverges to infinity if $R_{i}\in \mathrm{MSS}$.
Thus, $M_{i,T}$ can be employed for a \emph{pointwise} statistical inference
on whether $R_{i}$ belongs to the \textrm{MSS}. However, since our objective
is on estimation and inference of the $\mathrm{MSS}$, which is a set
potentially containing multiple assets, a \emph{uniformly} inference
procedure based on $M_{i,T}$ over $i=1,\ldots ,d$ is required.

Two technical challenges arise in conducting uniform inference. First, the
(asymptotic) joint distribution of $M_{i,T}$ for $i=1,\ldots ,d$ depends on
unknown parameters, making it non-pivotal. To address this, we propose a
resampling method based on the moving blocks bootstrap (MBB) 
\citep{kunsch1989,
liu1992, fitzenberger1998} to approximate the finite-sample\
\textquotedblleft null\textquotedblright\ joint distribution of $M_{i,T}$
for $i=1,\ldots ,d$. The MBB also accounts for potential serial correlation
in financial returns. Second, $M_{i,T}$ diverges to infinity with $T$ if and
only if $i\in \mathrm{MSS}$. To ensure the inference procedure is not
conservative and maintains exact control of size (Type-I error), the desired
\textquotedblleft null\textquotedblright\ joint distribution of $M_{i,T}$
should be concentrated on\ $i\notin \mathrm{MSS}$. However, since the $%
\mathrm{MSS}$ is unknown, this desired \textquotedblleft
null\textquotedblright\ joint distribution remains infeasible even with the
MBB. To address this issue, we adopt a step-down approach from the multiple
hypothesis testing literature (see, e.g., \cite{romano2005exact}). This
method iteratively refines the bootstrap critical value used for $\mathrm{MSS%
}$ estimation, improving the power of our procedure in identifying important
assets in $R$.

Our estimator of the $\mathrm{MSS}$ is formally defined as the subset of
assets whose $M_{i,T}$ exceeds the refined bootstrap critical\ value
obtained through the step-down approach. Additionally, the magnitude of $%
M_{i,T}$\ serves as a metric for evaluating the relative importance of the
assets and ranking them in $R$. We theoretically demonstrate that this $%
\mathrm{MSS}$ estimator covers the true $\mathrm{MSS}$ with probability
approaching 1 (wpa1), and converges to the exact $\mathrm{MSS}$ with
probability reaching any pre-specified level, such as 0.95 or 0.99.
Additionally, this estimator can be made consistent by letting the
pre-specified level approach 1 with increasing sample size.

As a by-product, our $\mathrm{MSS}$ estimation procedure can also be applied
to the conventional spanning test problem, offering more insights than
traditional spanning tests. Given a pre-specified benchmark asset set and an
additional asset set, our method can identify and estimate the $\mathrm{MSS}$
of all assets under consideration. To determine whether the benchmark set
spans the additional set, we can simply check whether the estimated $\mathrm{%
MSS}$ is a subset of the benchmark set. More importantly, by analyzing the
intersections of the estimated $\mathrm{MSS}$ with the benchmark set and the
additional set, we can identify which assets in the benchmark set become
redundant upon including the additional set, and which assets in the
additional set are truly valuable. This approach provides a more nuanced and
refined assessment of asset relevance, surpassing the binary conclusions of
conventional spanning tests.

The finite-sample performance of our proposed $\mathrm{MSS}$ estimation
procedure is assessed through extensive Monte Carlo simulations. We simulate
the data using a model with an autoregressive (AR) conditional mean and a
generalized autoregressive conditional heteroskedasticity (GARCH)
conditional variance, which effectively captures key stylized features of
financial returns, including serial and cross-sectional correlation as well
as volatility clustering. The simulation results demonstrate that the
empirical probability of the estimated $\mathrm{MSS}$ containing the true $%
\mathrm{MSS}$ approaches one as the sample size increases. Furthermore, the
empirical probability of the estimated $\mathrm{MSS}$ being identical to the
true $\mathrm{MSS}$ aligns closely with the nominal significance level for
sufficiently large sample sizes. These findings are consistent with the
asymptotic theory established for our method, demonstrating its robust
performance in finite samples.

We apply the proposed method to study the relative importance of stock
momentum factors and factor momentum strategies, along with a set of
well-established stock return factors.\footnote{%
We thank \cite{SJ2022} for kindly making their data available.}\ The main
findings from our empirical analysis are as follows. First, when either
individual stock momentum factor or factor momentum is combined with the
return factors, they are consistently included in the estimated $\mathrm{MSS}
$, highlighting the significance of return momentum in mean-variance
analysis. Second, when factor momentum coexists with all individual stock
momentum factors, it is consistently selected in the $\mathrm{MSS}$. At the
same time, individual stock momentum factors---such as the standard momentum
and the industry-adjusted momentum---also contribute to enhancing
mean-variance efficiency. Third, our empirical analysis reveals differing
relative importance between the two factor momentum strategies. When both
factor momentum strategies are included with other individual stock momentum
factors and return factors, only the momentum in the first ten principal
component factors is selected in the estimated $\mathrm{MSS}$. This result
aligns with \cite{SJ2022}, which suggests that factor momentum effectively
prices individual stock momentum and is generally concentrated in
high-eigenvalue principal components.\ Additionally, our method underscores
the importance of several individual stock momentum strategies, such as
standard momentum and industry-adjusted momentum, as well as other prominent
stock return factors, including excess market return, size, and betting
against beta.

Our study makes a direct contribution to the growing literature on
conventional spanning tests. \cite{huberman1987} derives the key conditions
under which a given set of assets spans the mean-variance frontier of a
larger set when additional assets are included, and introduces a likelihood
ratio test to assess the redundancy of the additional set of assets.\
Subsequent advancements in this field have been made by \cite{HJ1991}, \cite%
{ferson-foerster-keim-93}, \cite{desantis-93}, \cite{bekaert-urias-96}, \cite%
{deroon-nijman-werker-01}, \cite{amengual2010}, \cite{kan2012tests}, \cite%
{penaranda2012}, among others. However, as emphasized earlier, our study is
the first to focus on the identification and estimation of the $\mathrm{MSS}$%
, marking a significant departure from the existing literature on spanning
tests. Empirically, our work adds to the ongoing discussions on the
interplay between factor momentum and momentum factors, as detailed by \cite%
{GK2018}, \cite{SJ2022}, \cite{YY2023}, and \cite{AKL2023}. By
characterizing the $\mathrm{MSS}$ for various momentum strategies,
evaluating their relative importance, and ranking them within a large set of
assets, our approach provides novel insights into the interactions between
these factors and their role in mean-variance analysis.

The remainder of this paper is organized as follows. Section \ref{sec:t}
introduces the identification conditions for the \textrm{MSS} and details
the implementation of the proposed estimation and inference method.\ It also
establishes the method's asymptotic properties and demonstrates how slight
modifications can extend its applicability to other important problems in
empirical finance. Section \ref{sec:mc} presents simulation studies to
assess the finite-sample performance of the method, while Section \ref%
{sec:emp} offers an empirical application. Finally, Section \ref%
{sec:conclusion} concludes the paper. The Appendix includes proofs of the
main theoretical results, auxiliary lemmas used in these proofs, and
additional simulation results. The Supplemental Appendix contains detailed
proofs of the auxiliary lemmas.

\textit{Notation.} We use $a\equiv b$ to indicate that $a$ is defined as $b$%
.\ For any positive integer $m$, let $I_{m}$ denote the $m\times m$ identity
matrix. For any positive integers $m_{1}$ and $m_{2}$, $\mathbf{1}%
_{m_{1}\times m_{2}}$ and $\mathbf{0}_{m_{1}\times m_{2}}$ denote the $%
m_{1}\times m_{2}$ matrices of ones and zeros, respectively. For real
numbers $a_{1},\ldots ,a_{m}$, let $(a_{i})_{i\leq m}\equiv (a_{1},\ldots
,a_{m})^{\top }$, and let $a_{-i}$ denote the subvector of $(a_{i})_{i\leq m}
$ with $a_{i}$ excluded. The $\ell _{\infty }$ norm of $(a_{i})_{i\leq m}$
is given by\ $|(a_{i})_{i\leq m}|_{\infty }=\max_{i\leq m}\left\vert
a_{i}\right\vert $. Define the support of $(a_{i})_{i\leq m}$ as $\mathrm{%
Supp}_{(a_{i})_{i\leq m}}\equiv \{i=1,\ldots ,m:a_{i}\neq 0\}$. For any
matrices $A$ and $B$, $\mathrm{diag}(A,B)$ represents a block diagonal
matrix with $A$ and $B$ as its diagonal blocks, and $A\otimes B$ denotes the
Kronecker product of $A$ and $B$. Additionally, $A_{j,.}$ represents the $j$%
th row of the matrix $A$. For any positive integer $d$, let $\mathcal{M}%
_{d}\equiv \{1,\ldots ,d\}$, and for any positive integer $i\leq d$, let $%
\ell _{d,i}$ denote the $d\times 1$ vector whose $i$th entry is $1$, with
all other entries equal to $0$. For two sequences of positive numbers $a_{n}$
and $b_{n}$, we write $a_{n}\succ b_{n}$ if $a_{n}\geq c_{n}b_{n}$ for some
strictly positive sequence $c_{n}\rightarrow \infty $.

\setcounter{equation}{0} \renewcommand{\theequation}{\thesection.%
\arabic{equation}}


\section{Main Theory \label{sec:t}}

This section presents the main theoretical results of the paper. In
subsection \ref{subsec:id}, we examine the existence and uniqueness of the $%
\mathrm{MSS}$ and establish its identification condition. The identification
condition is constructive, as it is employed in subsection \ref{subsec:est}
to develop valid estimation and inference method for the $\mathrm{MSS}$.
Subsection\  \ref{subsec:other} demonstrates how the $\mathrm{MSS}$
estimation and inference procedure can be adapted for other important
applications in empirical finance.

\subsection{Identification Condition \label{subsec:id}}

For a given set of assets, represented by their returns $R\equiv
(R_{i})_{i\leq d}$, our goal is to identify the smallest subset, referred to
as the $\mathrm{MSS}$, such that the assets in this subset span the
mean-variance frontier of the original set. The $\mathrm{MSS}$ must satisfy
two key conditions: first, it must span the mean-variance frontier of the
original set; second, it cannot be further reduced, meaning that any proper
subset of the $\mathrm{MSS}$ cannot span the mean-variance frontier of the
original set. Based on these properties, we provide the formal definition of
the $\mathrm{MSS}$ below.

\begin{definition}
A subvector of $R$ is called a minimum spanning set ($\mathrm{MSS}$), if it
is the smallest subvector of $R$ which spans the mean-variance frontier of $%
R $.
\end{definition}

Since the mean-variance frontier of $R$ depends only on its mean $\mu $ and
variance $\Sigma $, which are consistently estimable, we assume in this
subsection that both $\mu $ and $\Sigma $ are known for the purpose of
investigating the identification condition of the $\mathrm{MSS}$. Given that
the set of assets $R$ spans its own mean-variance frontier, the $\mathrm{MSS}
$ is guaranteed to exist. Specifically, since there are $2^{d}-2$ nonempty
and proper subvectors of $R$, we can examine each of these subvectors and
identify those that span the mean-variance frontier of $R$. The $\mathrm{MSS}
$ will be the subvector(s) in this collection with the smallest dimension.
If no proper subvector of $R$ spans the mean-variance frontier of $R$, then
the $\mathrm{MSS}$ is $R$ itself.

A question of uniqueness naturally arises from the above discussion on the
existence of the $\mathrm{MSS}$: are there two or more distinct subvectors,
say $R_{K,1}$ and $R_{K,2}$ of $R$ with the same dimension, that span the
mean-variance frontier of $R$? The answer is no, as demonstrated in the
lemma below. This lemma also provides a constructive approach for
identifying the $\mathrm{MSS}$, which serves as the basis of our proposed
estimation and inference procedure.

\begin{lemma}
\label{ID}\ Suppose that $\Sigma $ is finite and non-singular.\ Then the $%
\mathrm{MSS}$ exists and is unique. Moreover, for any asset $R_{i}$ in $R$,\
consider the\ least squares (LS) regression of $R_{i}$ on the remaining
assets in $R$, denoted as\ $R_{-i}$: 
\begin{equation}
R_{i}=\alpha _{i}+\beta _{i}^{\top }R_{-i}+\varepsilon _{i}.  \label{ID_1}
\end{equation}%
Then the $\mathrm{MSS}$ satisfies:%
\begin{equation}
\left \{ 
\begin{array}{cc}
\alpha _{i}^{2}+(1_{d-1}^{\top }\beta _{i}-1)^{2}\neq 0, & \text{for any }%
i\in \mathrm{MSS} \\ 
\alpha _{i}^{2}+(1_{d-1}^{\top }\beta _{i}-1)^{2}=0, & \text{for any }%
i\notin \mathrm{MSS}%
\end{array}%
\right. .  \label{ID_2}
\end{equation}
\end{lemma}

Lemma \ref{ID}\ has three important implications. First, each asset\ $R_{i}$
in $R$ can be characterized by a pair of values $\theta _{i}\equiv \left(
\alpha _{i},1_{d-1}^{\top }\beta _{i}\right) ^{\top }$, which are uniquely
determined by $\mu $ and $\Sigma $. This pair, along with the condition in\ (%
\ref{ID_2}), enables us to obtain the\ $\mathrm{MSS}$ as follows:%
\begin{equation}
\mathrm{MSS}=\left \{ i\in \mathcal{M}_{d}\text{: }\alpha
_{i}^{2}+(1_{d-1}^{\top }\beta _{i}-1)^{2}>0\right \} .  \label{MSS_ID}
\end{equation}%
Second, the identification condition presented in Lemma \ref{ID} is
constructive and does not require any prior knowledge of the $\mathrm{MSS}$.
Given the population values $\mu $ and $\Sigma $, we only need to compute $%
\theta _{i}$ for $i\in \mathcal{M}_{d}$, and use (\ref{MSS_ID}) to determine
the $\mathrm{MSS}$. Third, Lemma \ref{ID} strengthens a key result from \cite%
{huberman1987}, which is widely used in the literature to assess the
redundancy of additional assets relative to a benchmark asset set.
Specifically,\ Proposition 3 of \cite{huberman1987} implies that $R_{-i}$
spans the mean-variance frontier of $R$ (and hence $R_{i}$ is redundant
relative to $R_{-i}$) if and only if:%
\begin{equation}
\alpha _{i}^{2}+(1_{d-1}^{\top }\beta _{i}-1)^{2}=0.  \label{ID_3}
\end{equation}%
While this result identifies the redundancy of an individual asset relative
to its complement in the full set, it does not provide a method for
determining the $\mathrm{MSS}$. Lemma \ref{ID} extends this finding by
showing that removing all assets satisfying (\ref{ID_3}) from $R$ yields the 
$\mathrm{MSS}$.

In practice, the value of $\theta _{i}$ for each asset $R_{i}$ is unknown
but can be estimated through LS regression of $R_{i}$ on $R_{-i}$. Combined
with the identification conditions established in Lemma \ref{ID}, this
enables the estimation and inference of the $\mathrm{MSS}$ in finite
samples. In the next subsection, we analyze the asymptotic properties of the
LS estimators of $\theta _{i}$ \emph{uniformly} over $i\in \mathcal{M}_{d}$%
.\ These properties facilitate the construction of $\mathrm{MSS}$ estimator
that cover the true $\mathrm{MSS}$ with probability approaching 1 (wpa1),
and asymptotically identify it with any desired level of confidence.

\subsection{Estimation and Inference of the Minimum Spanning Set \label%
{subsec:est}}

We first introduce some notations to simplify the definition of the
estimator for $\theta _{i}$, as well as the estimation procedure for the $%
\mathrm{MSS}$.\ The key value $\theta _{i}$ can be expressed as a linear
transformation of $\alpha _{i}$ and $\beta _{i}$: $\theta _{i}\equiv A\cdot
(\alpha _{i},\beta _{i}^{\top })^{\top }$ where $A\equiv \mathrm{diag}%
(1,1_{d-1}^{\top })$.\ For any observation $R_{t}\equiv (R_{i,t})_{i\leq d}$
where $t=1,\ldots T$, we let $\tilde{R}_{-i,t}\equiv \lbrack
1,R_{-i,t}^{\top }]^{\top }$ denote the regressors in the linear regression
specified in (\ref{ID_1}), $\hat{Q}_{-i}\equiv T^{-1}\sum_{t\leq T}\tilde{R}%
_{-i,t}\tilde{R}_{-i,t}^{\top }$ and $Q_{-i}\equiv \mathrm{E}[\tilde{R}%
_{-i,t}\tilde{R}_{-i,t}^{\top }]$ for any $i\in \mathcal{M}_{d}$.

Next, we describe the estimation procedures for the $\mathrm{MSS}$ and their
intuition. For each asset $R_{i}$, the LS estimator of $(\alpha _{i},\beta
_{i}^{\top })^{\top }$ is defined as:%
\begin{equation}
(\hat{\alpha}_{i},\hat{\beta}_{i}^{\top })^{\top }\equiv \hat{Q}%
_{-i}^{-1}T^{-1}\sum_{t\leq T}\tilde{R}_{-i,t}R_{i,t}.  \label{OLS}
\end{equation}%
Given the definition of $\theta _{i}$, we can estimate it with $\hat{\theta}%
_{i}\equiv A\cdot (\hat{\alpha}_{i},\hat{\beta}_{i}^{\top })^{\top }$. Using
the expression for $R_{i}$ in (\ref{ID_1}), the definition of $\hat{\theta}%
_{i}$ and the LS estimators$\ $in (\ref{OLS}), we obtain an expression for
the estimation error in $\hat{\theta}_{i}$:%
\begin{equation}
T^{1/2}(\hat{\theta}_{i}-\theta _{i})=A\hat{Q}_{-i}^{-1}\left(
T^{-1/2}\sum_{t\leq T}\tilde{R}_{-i,t}\varepsilon _{i,t}\right) .
\label{OLS_Exp}
\end{equation}%
Let\ $\Omega _{d,T,i}\equiv \mathrm{Var}(T^{-1/2}\sum_{t\leq T}\tilde{R}%
_{-i,t}\varepsilon _{i,t})$. For any \emph{fixed} $i$, we can use this
expression, and apply the law of large numbers (LLN) and the central limit
theorem (CLT) to show that $T^{1/2}(\hat{\theta}_{i}-\theta _{i})$ is
approximately distributed as normal with mean zero and variance $%
AQ_{-i}^{-1}\Omega _{d,T,i}Q_{-i}^{-1}A^{\top }$, denoted as 
\begin{equation*}
T^{1/2}(\hat{\theta}_{i}-\theta _{i})\overset{d}{\approx }N(0,\text{ }%
AQ_{-i}^{-1}\Omega _{d,T,i}Q_{-i}^{-1}A^{\top }).
\end{equation*}%
This pointwise result can be used to test whether a given asset $R_{i}$
belongs to the $\mathrm{MSS}$ or not. However, our goal is to estimate and
conduct statistical inference on the $\mathrm{MSS}$, which may include
multiple assets. Therefore, to ensure accurate estimation of the $\mathrm{MSS%
}$ and proper control of statistical inference errors, we need to conduct a
joint statistical inference on condition (\ref{ID_2}) for $i\in \mathcal{M}%
_{d}$, which requires approximating the finite-sample distribution of $\hat{%
\theta}_{i}$ \textit{uniformly} over $i\in \mathcal{M}_{d}$.

For the purpose of joint inference, we stack the expression in (\ref{OLS_Exp}%
) for different $i$ to obtain a joint representation of the estimation
errors for $\hat{\theta}$:%
\begin{equation}
T^{1/2}(\hat{\theta}-\theta )=\left( A\hat{Q}_{-i}^{-1}T^{-1/2}\sum_{t\leq T}%
\tilde{R}_{-i,t}\varepsilon _{i,t}\right) _{i\leq d},  \label{OLS_J_Exp}
\end{equation}%
where $\hat{\theta}\equiv (\hat{\theta}_{i})_{i\leq d}$ and $\theta \equiv
(\theta _{i})_{i\leq d}$. By the (uniform) consistency of $\hat{Q}_{-i}$,
the term on the right hand side of (\ref{OLS_J_Exp}) can be approximated by\ 
$T^{-1/2}\sum_{t\leq T}e_{t}$, where $e_{t}\equiv (e_{i,t})_{i\leq d}$ and $%
e_{i,t}\equiv AQ_{-i}^{-1}\tilde{R}_{-i,t}\varepsilon _{i,t}$.\ Intuitively,
by a CLT-type of argument, the finite sample distribution of\ $%
T^{-1/2}\sum_{t\leq T}e_{t}$ can be approximated by $\Omega _{d, T}^{1/2}%
\mathcal{N}_{d}$, where $\Omega _{d,T}\equiv \mathrm{Var}(T^{-1/2}\sum_{t%
\leq T}e_{t})$ and $\mathcal{N}_{d}$ denotes a standard normal random
vector.\ Therefore, the finite-sample distribution of $T^{1/2}(\hat{\theta}%
-\theta )$ can be approximated by the distribution of $\Omega _{d,T}^{1/2}%
\mathcal{N}_{d}$, denoted as 
\begin{equation}
T^{1/2}(\hat{\theta}-\theta )\overset{d}{\approx }\Omega _{d,T}^{1/2}%
\mathcal{N}_{d}.  \label{Coupling}
\end{equation}%
This intuition is employed to obtain critical values in our procedure for
estimating the $\mathrm{MSS}$.

We are now ready to introduce the test statistic and critical value used in
our estimation procedure. Let\ $A_{j,\cdot }$\ denote the $j$th row of $A$
for $j=1,2$. For each asset $R_{i}$, we construct%
\begin{equation}
M_{i,T}\equiv \max \left \{ \frac{T^{1/2}\left \vert \hat{\alpha}%
_{i}\right \vert }{\hat{s}_{i,1}},\text{ \ }\frac{T^{1/2}\left \vert
1_{d-1}^{\top }\hat{\beta}_{i}-1\right \vert }{\hat{s}_{i,2}}\right \} ,
\label{Max_i}
\end{equation}%
where\ $\hat{s}_{i,j}^{2}\equiv \hat{\sigma}_{\varepsilon
_{i}}^{2}A_{j,\cdot }\hat{Q}_{-i}^{-1}A_{j,\cdot }^{\top }$ and $\hat{\sigma}%
_{\varepsilon _{i}}^{2}\equiv T^{-1}\sum_{t\leq T}(R_{i,t}-\hat{\theta}%
_{i}^{\top }\tilde{R}_{-i,t})^{2}$ for $j=1,2$. Clearly, $M_{i,T}$ is the
maximum of the t-ratios for testing $\alpha _{i}=0$ and $1_{d-1}^{\top
}\beta _{i}=1$, respectively. Given the identification condition in (\ref%
{ID_2}) and the approximation result in (\ref{Coupling}), it follows that
for any $i\in \mathrm{MSS}$, $M_{i,T}\ $diverges as the sample size $T$
increases, while for any $i\notin \mathrm{MSS}$,\ $M_{i,T}$ can be
approximated in distribution by 
\begin{equation}
\tilde{M}_{i,T}\equiv \left \vert (\ell _{d,i}^{\top }\otimes \mathrm{diag}%
(s_{i,1}^{-1},s_{i,2}^{-1}))\Omega _{d,T}^{1/2}\mathcal{N}_{d}\right \vert
_{\infty },  \label{M_tilda}
\end{equation}%
where $s_{i,j}^{2}\equiv \sigma _{\varepsilon _{i}}^{2}A_{j,\cdot
}Q_{-i}^{-1}A_{j,\cdot }^{\top }$ and $\sigma _{\varepsilon _{i}}^{2}\equiv 
\mathrm{E}[\varepsilon _{i,t}^{2}]$. Since the assets in the $\mathrm{MSS}$
tend to have larger $M_{i,T}$ values than those that are not in the $\mathrm{%
MSS}$, a formal statistical inference procedure should provide a critical
value under a pre-specified significance level $p\in (0,1)$ to determine
when an asset with a large $M_{i,T}$ value can be included in the $\mathrm{%
MSS}$.

Ideally, for any small $p\in (0,1)$, the critical value should depend only
on the assets that are not in the $\mathrm{MSS}$. This is because, in view
of Lemma \ref{ID}, the\  \textquotedblleft null hypothesis\textquotedblright
: 
\begin{equation*}
\alpha _{i}^{2}+(1_{d-1}^{\top }\beta _{i}-1)^{2}=0
\end{equation*}%
holds only for $i\notin \mathrm{MSS}$. Therefore, if the $\mathrm{MSS}$ is a
proper subset of $\mathcal{M}_{d}$, we would use the\ $(1-p)$-quantile of $%
\max_{i\notin \mathrm{MSS}}\tilde{M}_{i,T}$, denoted as $cv_{1-p}^{u}$, as
the critical value. This leads to an infeasible estimator of $\mathrm{MSS}$: 
\begin{equation}
\widehat{\mathrm{MSS}}_{p}^{u}\equiv \left \{ i\in \mathcal{M}_{d}\text{: }%
M_{i,T}>cv_{1-p}^{u}\right \} .  \label{inf_MSS_est_2}
\end{equation}%
Under certain mild conditions, such as the sufficient conditions stated in
Theorem \ref{T2} below, $\widehat{\mathrm{MSS}}_{p}^{u}$ possesses the
desirable properties of covering the true $\mathrm{MSS}$ wpa1, and
overestimates it with a small probability $p$.

However, the critical value $cv_{1-p}^{u}$ is not practical to use for two
main reasons: (i) the distribution of $(\tilde{M}_{i,T})_{i\leq d}$ is
unknown due to the presence of nuisance parameters $(s_{i,j})_{j\leq 2}$ ($%
i\leq d$) and $\Omega _{d,T}$; and (ii) there is no prior knowledge about
the\ $\mathrm{MSS}$. The first challenge can be addressed by estimating the
distribution of $(\tilde{M}_{i,T})_{i\leq d}$ through either plugging in
consistent estimators of the nuisance parameters $(s_{i,j})_{j\leq 2}$\ ($%
i\leq d$) and $\Omega _{d,T}$ directly, or by using resampling methods such
as the bootstrap. In this paper, we employ the moving blocks 
\citep{kunsch1989, liu1992,
fitzenberger1998}\ bootstrap (MBB) to approximate the distribution of $(%
\tilde{M}_{i,T})_{i\leq d}$. The second challenge is more delicate. One
solution is to use a known upper bound of $\max_{i\notin \mathrm{MSS}}\tilde{%
M}_{i,T}$, e.g., $\max_{i\leq d}\tilde{M}_{i,T}$, to obtain a critical value
which is larger than $cv_{1-p}^{u}$. While straightforward to implement,
this method is conservative, potentially leading to an underestimation of
the $\mathrm{MSS}$ in finite samples.\ Instead, we adopt the step-down
procedure (see, e.g., \cite{romano2005exact}),\ which iteratively refines
the bootstrap critical value, thereby enhancing the power of our procedure
in estimating the $\mathrm{MSS}$.

Details of our $\mathrm{MSS}$ estimation procedure are provided in the
algorithm below.

\bigskip

\noindent \textbf{Algorithm: A Bootstrap }$\mathbf{MSS}$\textbf{\ Estimation
Procedure}

\noindent Step \textbf{1}.\ For each asset $R_{i}$, run the linear
regression specified in (\ref{ID_1}) to obtain the estimators of $\hat{\alpha%
}_{i}$ and $1_{d-1}^{\top }\hat{\beta}_{i}^{\top }$, and calculate $M_{i,T}$
specified in (\ref{Max_i}).

\noindent Step \textbf{2}.\ Given a bandwidth\ $\ell $, define moving blocks 
$B_{j}=\{R_{j},\ldots ,R_{j+\ell -1}\}$ for $j=1,\ldots ,q$ where $q\equiv
T-\ell +1$.\ This will give $q$ blocks $\{B_{j}\}_{j\leq q}$.

\noindent Step \textbf{3}.\ Let $m\equiv \lbrack T/\ell ]$ and $T_{B}\equiv
m\ell $. Resample $m$ blocks $Z_{j}^{b}\equiv \{R_{\ell (j-1)+1}^{b},\ldots
,R_{\ell (j-1)+\ell }^{b}\}$ independently from\ $\{B_{j}\}_{j\leq q}$ to
obtain a bootstrap sample $\{R_{t}^{b}\}_{t\leq T_{B}}\equiv
\{Z_{j}^{b}\}_{j\leq m}$, and then calculate%
\begin{equation}
M_{i,T}^{b}\equiv \max \left \{ \frac{T^{1/2}\left \vert \hat{\alpha}%
_{i}^{b}-\hat{\alpha}_{i}\right \vert }{\hat{s}_{i,1}^{b}},\text{\ }\frac{%
T^{1/2}\left \vert 1_{d-1}^{\top }(\hat{\beta}_{i}^{b}-\hat{\beta}%
_{i})\right \vert }{\hat{s}_{i,2}^{b}}\right \} ,  \label{b_Max_i}
\end{equation}%
where $\hat{\alpha}_{i}^{b}$, $\hat{\beta}_{i}^{b}$ and $\hat{s}_{i,j}^{b}$\
denote the bootstrap counterparts of $\hat{\alpha}_{i}$, $\hat{\beta}_{i}$
and $\hat{s}_{i,j}$, respectively.

\noindent Step \textbf{4}.\ Repeat Step 3\ $B$ times to obtain $%
\{(M_{i,T}^{b})_{i\leq d}\}_{b\leq B}$.

\noindent Step \textbf{5}.\ For any $\mathcal{S}\subseteq \mathcal{M}_{d}$,
let $cv_{1-p,T}^{b}(\mathcal{S})$ denote the $(1-p)$-conditional quantile of 
$\max_{i\in \mathcal{S}}M_{i,T}^{b}$ given the data. Starting with $\widehat{%
\mathrm{MSS}}_{p,0}=\varnothing $, we update the estimate of the $\mathrm{MSS%
}$ through: 
\begin{equation}
\widehat{\mathrm{MSS}}_{p,j}=\left \{ i\in \mathcal{M}_{d}\text{: }%
M_{i,T}>cv_{1-p,T}^{b}(\mathcal{M}_{d}\backslash \widehat{\mathrm{MSS}}%
_{p,j-1})\right \} .  \label{updating_MSS}
\end{equation}%
The updating process stops at $j^{\ast }$ when $\widehat{\mathrm{MSS}}%
_{p,j^{\ast }}=\widehat{\mathrm{MSS}}_{p,j^{\ast }-1}$\ or\ $\widehat{%
\mathrm{MSS}}_{p,j^{\ast }}=\mathcal{M}_{d}$.

\noindent Step \textbf{6}. The final $\mathrm{MSS}$ estimate is then defined
as $\widehat{\mathrm{MSS}}_{p}=\widehat{\mathrm{MSS}}_{p,j^{\ast }}$.

\bigskip

Steps 2--4 in the algorithm above describe the MBB procedure for obtaining
the bootstrap statistics $(M_{i,T}^{b})_{i\leq d}$, while Step 5 represents
the step-down procedure for constructing the critical value used in our $%
\mathrm{MSS}$ estimation procedure. Our $\mathrm{MSS}$ estimator can be
formally defined as:%
\begin{equation}
\widehat{\mathrm{MSS}}_{p}\equiv \left \{ i\in \mathcal{M}_{d}\text{: }%
M_{i,T}>cv_{1-p,T}^{b}\right \} ,  \label{MSS_Est}
\end{equation}%
where $cv_{1-p,T}^{b}\equiv cv_{1-p,T}^{b}(\mathcal{M}_{d}\backslash 
\widehat{\mathrm{MSS}}_{p,j^{\ast }})$.\footnote{%
We set $cv_{1-p,T}^{b}$ to zero if $\widehat{\mathrm{MSS}}_{p,j^{\ast }}=%
\mathcal{M}$.} When the size of the $\mathrm{MSS}$ is small and the
identification condition in (\ref{ID_2})\ is nearly satisfied, the test
based on $cv_{1-p,T}^{b}\ $may cause the estimator $\widehat{\mathrm{MSS}}%
_{p}$ to be an empty set, particularly in small sample sizes. However, since
the $\mathrm{MSS}$ cannot be empty, we implement a finite-sample adjustment
by estimating it as the set of assets with the highest values of $M_{i,T}$
in such cases.\footnote{%
In the proof of Theorem \ref{T2}, we show that\ $\widehat{\mathrm{MSS}}_{p}$%
, as obtained from our estimation algorithm, is nonempty wpa1. Consequently,
the finite-sample adjustment becomes asymptotically negligible.}

\begin{theorem}
\label{T2} Suppose that\ Assumptions\  \ref{A1} and \ref{A2} in the Appendix
holds, and moreover%
\begin{equation}
\min_{i\in \mathrm{MSS}}\left( \alpha _{i}^{2}+\frac{(1_{d-1}^{\top }\beta
_{i}-1)^{2}}{d-1}\right) \succ T^{-1}.  \label{T2_1}
\end{equation}%
Then we have $\lim_{T\rightarrow \infty }\mathrm{P}(\mathrm{MSS}\subseteq 
\widehat{\mathrm{MSS}}_{p})=1$. Moreover, if $\mathrm{MSS}$ is a proper
subset of $\mathcal{M}_{d}$, then $\lim_{T\rightarrow \infty }\mathrm{P}(%
\mathrm{MSS}=\widehat{\mathrm{MSS}}_{p})=1-p$.
\end{theorem}

Theorem \ref{T2} establishes that the $\mathrm{MSS}$ estimator $\widehat{%
\mathrm{MSS}}_{p}$ retains the desirable properties of covering the true $%
\mathrm{MSS}$ wpa1, and overestimates the $\mathrm{MSS}$ with probability
exactly $p$ as the sample size $T$ increases. If $\mathrm{MSS}=\mathcal{M}%
_{d}$, then the first result of Theorem \ref{T2} implies that $%
\lim_{T\rightarrow \infty }\mathrm{P}(\mathrm{MSS}=\widehat{\mathrm{MSS}}%
_{p})=1$. The condition in (\ref{T2_1}) specifies the least favorable
scenario under which these desirable properties still hold. This condition
is derived from the identification conditions in (\ref{ID_2}), and is
similarly adjusted to resemble the local power analysis of statistical
hypothesis test. If there exists a constant $K>0$ such that (\ref{ID_2})
holds for any asset $R_{i}$ in $\mathrm{MSS}$ with%
\begin{equation*}
\alpha _{i}^{2}+(1_{d-1}^{\top }\beta _{i}-1)^{2}>K,
\end{equation*}%
then (\ref{T2_1}) is trivially satisfied. More importantly,\ (\ref{T2_1})\
allows for the identification conditions (\ref{ID_2}) to nearly hold, in the
sense that some of $\alpha _{i}^{2}+(1_{d-1}^{\top }\beta _{i}-1)^{2}$ ($%
i\in \mathrm{MSS}$) are close to zero.

From the definitions of $M_{i,T}$ and $\widehat{\mathrm{MSS}}_{p}$, we have: 
\begin{equation}
\widehat{\mathrm{MSS}}_{p}=\left \{ i\leq d\text{: }\left \vert \hat{\alpha}%
_{i}\right \vert >T^{-1/2}\hat{s}_{i,1}cv_{1-p,T}^{b}\right \} \bigcup \left
\{ i\leq d\text{:\ }|1_{d-1}^{\top }\hat{\beta}_{i}-1|>T^{-1/2}\hat{s}%
_{i,2}cv_{1-p,T}^{b}\right \} .  \label{MSS_UB}
\end{equation}%
Here, $[0,$ $T^{-1/2}\hat{s}_{i,1}cv_{1-p,T}^{b}]$ and $[0$, $T^{-1/2}\hat{s}%
_{i,2}cv_{1-p,T}^{b}]$\ can be interpreted as\ uniform confidence bands for
estimating the $\mathrm{MSS}$ through testing $\alpha _{i}=0$ and $%
1_{d-1}^{\top }\beta _{i}-1=0$ over $i\in \mathcal{M}_{d}$. From the
expression in (\ref{MSS_UB}), it follows that an asset $i$ is selected into $%
\widehat{\mathrm{MSS}}_{p}$, if either $\left \vert \hat{\alpha}_{i}\right
\vert $, $|1_{d-1}^{\top }\hat{\beta}_{i}-1|$, or both, exceed their
respective upper bound of the uniform confidence band.\footnote{%
As noted in \cite{kan2012tests}, these two cases, i.e., $|\alpha _{i}|>0$
and $|1_{d-1}^{\top }\beta _{i}-1|>0$, are subject to explicit economic
interpretation. Specifically, $|\alpha _{i}|>0$ implies asset $i$
contributes to the tangency portfolio, while $|1_{d-1}^{\top }\beta
_{i}-1|>0 $ corresponds to the scenario where asset $i$ contributes to the
global minimum variance portfolio. For more details on the derivations, see
Section 2.1 in \cite{kan2012tests}.}\ Therefore, our method provides
detailed insights into the sources of contribution for the selected assets.
Additionally, by ranking all assets in the full set according to their $%
M_{i,T}$ values, we can directly measure their relative importance.\footnote{%
For an empirical illustration of these ideas, refer to Figures \ref%
{fig:diag_MSS_2} and \ref{fig:diag_MSS}, as well as the related discussion
in Section \ref{sec:emp}.}

\subsection{Other Applications \label{subsec:other}}

Although the procedure in the previous section is primarily designed for
inference on the $\mathrm{MSS}$, it can be adapted to conduct statistical
inference for other unknown parameters of practical interest. In this
subsection, we present four illustrative examples.

First, in the proof of Lemma \ref{ID} in the Appendix, we show that the $%
\mathrm{MSS}$ is equivalent to the union of the supports of the tangency
portfolio and the global minimum variance portfolio, which correspond to the
supports of $(\alpha _{i})_{i\leq d}$ and $(1_{d-1}^{\top }\beta
_{i}-1)_{i\leq d}$, respectively. Accordingly, our inference procedure for
the $\mathrm{MSS}$ can be slightly modified to perform inference on the
minimal asset sets required for constructing these two important portfolios.

We next describe the modifications needed to adapt the $\mathrm{MSS}$
inference procedures for the tangency portfolio support,\ defined as\ $%
\mathrm{TAN}\equiv \{i\in \mathcal{M}_{d}:\alpha _{i}\neq 0\}$. For $i\in 
\mathcal{M}_{d}$, we define%
\begin{equation}
M_{i,T}(\alpha )\equiv \frac{T^{1/2}\left \vert \hat{\alpha}_{i}\right \vert 
}{\hat{s}_{i,1}}\text{ \  \  \ and \  \  \ }M_{i,T}^{b}(\alpha )\equiv \frac{%
T^{1/2}\left \vert \hat{\alpha}_{i}^{b}-\hat{\alpha}_{i}\right \vert }{\hat{s%
}_{i,1}^{b}}.  \label{M_a&boot}
\end{equation}%
By applying the algorithm from the previous section to estimate the $\mathrm{%
MSS}$, but replacing $M_{i,T}$ and $M_{i,T}^{b}$ by $M_{i,T}(\alpha )$ and $%
M_{i,T}^{b}(\alpha )$, respectively, we obtain the bootstrap critical value $%
cv_{1-p,T}^{b}(\alpha )$. This leads to the following estimator for $\mathrm{%
TAN}$:%
\begin{equation}
\widehat{\mathrm{TAN}}_{p}\equiv \left \{ i\in \mathcal{M}_{d}\text{: }%
M_{i,T}(\alpha )>cv_{1-p,T}^{b}(\alpha )\right \} .  \label{2nd_Tan_Est}
\end{equation}%
Using the arguments presented in the proof of Theorem \ref{T2}, it can be
shown that: $\widehat{\mathrm{TAN}}_{p}$ covers $\mathrm{TAN}$ with
probability approaching 1 and is identical to $\mathrm{TAN}$ with
probability converging to $1-p$.

The estimation procedure for the support of the global minimum variance
portfolio, defined as $\mathrm{GMV}\equiv \{i\in \mathcal{M}%
_{d}:1_{d-1}^{\top }\beta _{i}\neq 1\}$ is constructed analogously.
Specifically, by replacing $M_{i,T}(\alpha )$ and $M_{i,T}^{b}(\alpha )$ in (%
\ref{M_a&boot}) with 
\begin{equation}
M_{i,T}(\beta )\equiv \frac{T^{1/2}\left \vert 1_{d-1}^{\top }\hat{\beta}%
_{i}-1\right \vert }{\hat{s}_{i,2}}\text{ \ and \ }M_{i,T}^{b}(\beta )\equiv 
\frac{T^{1/2}\left \vert 1_{d-1}^{\top }(\hat{\beta}_{i}^{b}-\hat{\beta}%
_{i})\right \vert }{\hat{s}_{i,2}^{b}},  \label{M_b&boot}
\end{equation}%
for $i\in \mathcal{M}_{d}$, one can derive the corresponding
estimation/inference procedure for $\mathrm{GMV}$. For brevity, we omit the
detailed exposition.

In practice, researchers may also be interested in identifying the subset
of\ $\mathrm{TAN}$ consisting of assets with positive $\alpha _{i}$. We
denote this set as $\mathrm{TAN}^{+}\equiv \{i\in \mathcal{M}_{d}:\alpha
_{i}>0\}$. To obtain estimation/inference procedures for this set, we define%
\begin{equation}
\tilde{M}_{i,T}(\alpha )\equiv \frac{T^{1/2}\hat{\alpha}_{i}}{\hat{s}_{i,1}}%
\text{ \  \  \ and \  \  \ }\tilde{M}_{i,T}^{b}(\alpha )\equiv \frac{T^{1/2}(%
\hat{\alpha}_{i}^{b}-\hat{\alpha}_{i})}{\hat{s}_{i,1}^{b}}.
\label{M_a_+&boot}
\end{equation}%
Applying the algorithm from the previous section to estimate the $\mathrm{MSS%
}$, but replacing $M_{i,T}$ and $M_{i,T}^{b}$ by $\tilde{M}_{i,T}(\alpha )$
and $\tilde{M}_{i,T}^{b}(\alpha )$, respectively, we obtain the bootstrap
critical value $cv_{1-p,T}^{+,b}(\alpha )$. This leads to the following
estimator for $\mathrm{TAN}^{+}$:%
\begin{equation*}
\widehat{\mathrm{TAN}}_{p}^{+}\equiv \left \{ i\in \mathcal{M}_{d}\text{: }%
\tilde{M}_{i,T}(\alpha )>cv_{1-p,T}^{+,b}(\alpha )\right \} .
\end{equation*}%
Similar results to those stated in Theorem \ref{T2} can be established for $%
\widehat{\mathrm{TAN}}_{p}^{+}$ for valid inference on $\mathrm{TAN}^{+}$.

Finally, our procedure can be modified to perform joint inference on the
redundancy of an additional set of assets, denoted as $R_{N}\equiv
(R_{i})_{i\in \mathcal{N}}$ relative to a benchmark set of assets,\ $%
R_{K}\equiv (R_{i})_{i\in \mathcal{K}}$, where $\mathcal{N}$ and $\mathcal{K}
$ form a mutually exclusive and exhaustive partition of $\mathcal{M}_{d}$.
For each asset $R_{i}$ in $R_{N}$,\ we consider the following regression:%
\begin{equation*}
R_{i}=\alpha _{K,i}+\beta _{K,i}^{\top }R_{K}+\varepsilon _{K,i},
\end{equation*}%
from which we define the test statistic: 
\begin{equation}
M_{i,T}(\mathcal{N})\equiv \max \left \{ \frac{T^{1/2}\left \vert \hat{\alpha%
}_{K,i}\right \vert }{\hat{s}_{K,i,1}},\text{ \ }\frac{T^{1/2}\left \vert
1_{k}^{\top }\hat{\beta}_{K,i}-1\right \vert }{\hat{s}_{K,i,2}}\right \} ,
\label{M_N}
\end{equation}%
where\ $\hat{s}_{K,i,j}$ is constructed analogously to $\hat{s}_{i,j}$ ($%
j=1,2$). To determine the critical value for inference, we define the
bootstrap counterpart:%
\begin{equation}
M_{i,T}^{b}(\mathcal{N})\equiv \max \left \{ \frac{T^{1/2}\left \vert \hat{%
\alpha}_{K,i}^{b}-\hat{\alpha}_{K,i}\right \vert }{\hat{s}_{K,i,1}^{b}},%
\text{ \ }\frac{T^{1/2}\left \vert 1_{k}^{\top }(\hat{\beta}_{K,i}^{b}-\hat{%
\beta}_{K,i})\right \vert }{\hat{s}_{K,i,2}^{b}}\right \} ,  \label{M_N_Boot}
\end{equation}%
where $k$ is the cardinality of $\mathcal{K}$, $\hat{\alpha}_{K,i}^{b}$, $%
\hat{\beta}_{K,i}^{b}$, $\hat{s}_{K,i,1}^{b}$ and $\hat{s}_{K,i,2}^{b}$ are
the bootstrap versions of $\hat{\alpha}_{K,i}$, $\hat{\beta}_{K,i}$, $\hat{s}%
_{K,i,1}$ and $\hat{s}_{K,i,2}$, respectively.

Let $\mathcal{N}^{\ast }$ denote the subset of $\mathcal{N}$ consisting of
nonredundant assets, i.e., those that contribute to expanding the
mean-variance efficient frontier of $R_{K}$. We can estimate $\mathcal{N}%
^{\ast }$ by%
\begin{equation*}
\widehat{\mathcal{N}}_{p}^{\ast }\equiv \left \{ i\in \mathcal{N}\text{: \ }%
M_{i,T}(\mathcal{N})>cv_{1-p,T}^{b}(\mathcal{N})\right \} ,
\end{equation*}%
where\ $cv_{1-p,T}^{b}(\mathcal{N})$ denotes the bootstrap critical value
from the algorithm in the previous section with\ $M_{i,T}$ and $M_{i,T}^{b}$
replaced by $M_{i,T}(\mathcal{N})$ and $M_{i,T}^{b}(\mathcal{N})$,
respectively. The theoretical properties established for $\mathrm{MSS}$ in\
Theorem \ref{T2} also hold for $\widehat{\mathcal{N}}_{p}^{\ast }$.
Specifically, $\widehat{\mathcal{N}}_{p}^{\ast }$ covers $\mathcal{N}^{\ast
} $ wpa1, and is identical to $\mathcal{N}^{\ast }$ with probability
converging to $1-p$.

\setcounter{equation}{0} \renewcommand{\theequation}{\thesection.\arabic{equation}}




\section{Monte Carlo Simulations\label{sec:mc}}

In this section, we examine the finite-sample performance of the proposed $%
\mathrm{MSS}$ estimator using Monte Carlo simulations. The simulation design
is detailed in subsection \ref{sec-sim_dgp}, and the results are presented
in subsection \ref{sec-sim_rslt}.

\subsection{The Simulation Setting\label{sec-sim_dgp}}

We use a vector autoregressive (VAR) model to specify the conditional mean
and a GARCH model to capture the conditional variance of returns. This
VAR-GARCH framework effectively reproduces key stylized features of stock
returns, including serial and cross-sectional correlations as well as
volatility clustering.

Specifically, the returns are generated using the following equations: 
\begin{eqnarray}
R_{K,t} &=&\mu \cdot \mathbf{1}_{K}+\mathbf{A}R_{K,t-1}+\eta _{K,t},
\label{S_eq_1} \\
R_{N,t} &=&\mathbf{a}+\mathbf{B}R_{K,t}+\eta _{N,t}.  \label{S_eq_2}
\end{eqnarray}%
Here, (\ref{S_eq_1}) specifies a VAR of order one for the assets $R_{K,t}$
in the benchmark MSS, where\ $\mu \cdot \mathbf{1}_{K}$ and $\mathbf{A}$
denote the mean vector and the autoregressive coefficient matrix,
respectively. The returns of possible redundant assets $R_{N,t}$ are
connected to those in the benchmark $\mathrm{MSS}$ via (\ref{S_eq_2}), where 
$\mathbf{a}$ and $\mathbf{B}$ are parameters used to define the actual MSS
in accordance with Lemma \ref{ID}. To incorporate GARCH effects, we define $%
\eta _{t}\equiv (\eta _{K,t}^{\top },\eta _{N,t}^{\top })^{\top }$, where
the $i$th component of $\eta _{t}$, denoted as $\eta _{i,t}$ for $i=1,\ldots
,K+N$, satisfies: 
\begin{equation*}
\eta _{i,t}=d_{i,t}v_{i,t}\text{, \ where }d_{i,t}=(0.1+0.1\eta
_{i,t-1}^{2}+0.8d_{i,t-1}^{2})^{1/2},
\end{equation*}%
with $(v_{i,t})_{i\leq K+N}$ being a $(K+N)\times 1$ standard normal random
vector.

In the simulation, we consider $K+N=8$, where $K\in \{1,3,5,7\}$, and the
sample size $T\in \{120,180,240,300\}$, corresponding to monthly
observations spanning 10 to 25 years---periods commonly encountered in
practice. For the data-generating process of the MSS asset returns, we set $%
\mu =0$ and define the autoregressive matrix $\mathbf{A}\equiv
(a_{i,j})_{i,j\leq K}$ with $a_{i,j}=\rho ^{|i-j|+1}$\ and\ $\rho =0.1$. We
ensure the assets whose returns are governed by (\ref{S_eq_2}) are redundant
by setting $\mathbf{a}=\mathbf{0}_{N\times 1}$ and constraining the row sums
of $\mathbf{B}$ to equal 1. For simplicity, we specify $\mathbf{B}=K^{-1}%
\mathbf{1}_{N\times K}$. The block size $\ell $ in the MBB procedure is set
to $\lfloor 1.2T^{1/4}\rfloor $.

The infeasible estimator $\widehat{\mathrm{MSS}}_{p}^{u}$ and the proposed
estimator $\widehat{\mathrm{MSS}}_{p}$, as defined in (\ref{inf_MSS_est_2})
and (\ref{updating_MSS}) respectively, are investigated in the simulation
study. For both estimators, we set $p=0.05$, and evaluate their
finite-sample performance using 10,000 Monte Carlo replications.

\subsection{Simulation Results\label{sec-sim_rslt}}

Guided by theoretical insights, we first examine the empirical probabilities
of the estimated $\mathrm{MSS}$ containing the true $\mathrm{MSS}$, as
reported in the second and third columns of Table \ref{tab:sim1}. Consistent
with the main theory established in Section \ref{sec:t}, the simulation
results indicate that both the infeasible and proposed $\mathrm{MSS}$
estimators include the true $\mathrm{MSS}$ with high probability across all
combinations of $K$ and $N$, provided the sample size is moderately large
(e.g., $T\geq 180$), and this probability approaches 1 when the sample size
increases to $240$.

Next, we assess the accuracy of the estimated $\mathrm{MSS}$ in identifying
the true $\mathrm{MSS}$, measured as the probability that the $\mathrm{MSS}$
estimator exactly matches the true MSS. This property is particularly
important in practice, as it demonstrates the effectiveness of the $\mathrm{%
MSS}$ estimators in reducing the number of assets while maintaining the
mean-variance efficient frontier. The results, presented in the fourth and
fifth columns of Table \ref{tab:sim1}, show that the proposed MSS estimator $%
\widehat{\mathrm{MSS}}_{p}$ can identify the true $\mathrm{MSS}$ with high
probability when the sample size is sufficiently large, across all
combinations of $K$ and $N$ (e.g., $T\geq 180$). When the sample size $T$
reaches $240$, the probability of $\widehat{\mathrm{MSS}}_{p}$ coinciding
with the true $\mathrm{MSS}$ exceeds $0.9$ and converges to the nominal
value of $0.95$ as $T\ $increases to $300$.

Notably, the finite-sample performance of $\widehat{\mathrm{MSS}}_{p}$
closely matches that of the infeasible estimator it aims to mimic.
Specifically, when the number of benchmark assets $K$ is small (e.g., $K\leq
3$), the empirical probabilities that $\widehat{\mathrm{MSS}}_{p}$ either
contains or exactly matches the true $\mathrm{MSS}$ are nearly identical to
those of the infeasible $\mathrm{MSS}$ estimator. Performance differences
between these two estimators only arise when $K$ increases and the sample
size $T$ is relatively small (e.g., $T=120$). However, as $T$ grows
sufficiently large (e.g., $T\geq 240$),\ the proposed estimator $\widehat{%
\mathrm{MSS}}_{p}$ achieves performance comparable to the infeasible
estimator. These findings confirm that $\widehat{\mathrm{MSS}}_{p}$
effectively restores the ideal performance of the infeasible estimator in
finite samples.

In summary, the simulation results in Table \ref{tab:sim1} strongly align
with our theoretical findings: when the sample size is sufficiently large,
the estimated $\mathrm{MSS}$ not only contains but also coincides with the
true $\mathrm{MSS}$ with high probability. These results strongly support
the practical effectiveness of the proposed $\mathrm{MSS}$ estimator in
real-world applications.

\begin{table}[htbp]
\caption{Simulation Results for the MSS Estiamtors}
\label{tab:sim1}\centering  {\small \ 
\setlength \extrarowheight{-2.0pt}
		\begin{threeparttable} 
\begin{tabularx}{0.85\textwidth}{lcccccc}\toprule
				& & \multicolumn{2}{c}{$\mathrm{P}(\mathrm{MSS} \subset \widehat{\mathrm{MSS}})$} & & \multicolumn{2}{c}{$\mathrm{P}(\mathrm{MSS} = \widehat{\mathrm{MSS}})$} \\ \cline{3-4} \cline{6-7} 
				&    & Proposed Est. & Infeasible Est. &    & Proposed Est. & Infeasible Est.  \\
				\hline 
				& & \multicolumn{5}{c}{\textbf{Panel A}: $(K, N) = (1, 7)$} \\
				$T=120$ & &   0.939   &  0.937    &   &  0.898   &  0.904   \\
				$T=180$ & &   0.986   &  0.986    &   &  0.939   &  0.948   \\
				$T=240$ & &   0.998   &  0.998    &   &  0.943   &  0.953   \\
				$T=300$ & &   1.000   &  1.000    &   &  0.945   &  0.955   \\
				\hline
				& & \multicolumn{5}{c}{\textbf{Panel B}: $(K, N) = (3, 5)$} \\
				$T=120$ & &   0.922   &  0.922    &   &  0.877   &  0.887   \\
				$T=180$ & &   0.995   &  0.995    &   &  0.940   &  0.952   \\
				$T=240$ & &   1.000   &  1.000    &   &  0.941   &  0.953   \\
				$T=300$ & &   1.000   &  1.000    &   &  0.939   &  0.954   \\
				\hline	
				& & \multicolumn{5}{c}{\textbf{Panel C}: $(K, N) = (5, 3)$}  \\
				$T=120$ & &   0.828   &  0.825    &   &  0.769   &  0.790   \\
				$T=180$ & &   0.985   &  0.984    &   &  0.916   &  0.938   \\
				$T=240$ & &   0.999   &  0.998    &   &  0.927   &  0.952   \\
				$T=300$ & &   1.000   &  1.000    &   &  0.940   &  0.948   \\
				\hline	
				& & \multicolumn{5}{c}{\textbf{Panel D}: $(K, N) = (7, 1)$} \\
				$T=120$ & &   0.662   &  0.719    &   &  0.620   &  0.681   \\
				$T=180$ & &   0.953   &  0.965    &   &  0.898   &  0.915   \\
				$T=240$ & &   0.994   &  0.997    &   &  0.936   &  0.944   \\
				$T=300$ & &   0.999   &  0.999    &   &  0.947   &  0.950   \\
				\bottomrule
			\end{tabularx}
\begin{tablenotes}[flushleft]
				\item \textit{Note:} This table presents the simulation results for the MSS estimators described in Section \ref{sec:t}. The second and third columns display the empirical probabilities of the estimated MSS containing the true MSS, denoted as $\mathrm{P}(\mathrm{MSS} \subset \widehat{\mathrm{MSS}})$, for the proposed and infeasible MSS estimators. The fourth and fifth columns report the empirical probabilities of correctly identifying the true MSS, denoted as $\mathrm{P}(\mathrm{MSS} = \widehat{\mathrm{MSS}})$, for the proposed and infeasible MSS estimators, respectively. The significance level $p$ is set at 0.05. All empirical probabilities are calculated based on 10,000 Monte Carlo replications.
			\end{tablenotes}
\end{threeparttable} }
\end{table}

\setcounter{equation}{0} \renewcommand{\theequation}{\thesection.\arabic{equation}}


\section{An Empirical Application\label{sec:emp}}

Momentum, which posits that assets' past returns can effectively predict
their future returns, has been identified across various asset classes and
time periods. Due to its prevalence, momentum has been incorporated into
widely used asset pricing models to explain cross-sectional returns, such as
the Carhart four-factor model \citep{carhart1997}. Conversely, several
theories have been proposed to explain this anomaly, including time-varying
risk , behavioral biases, and trading frictions; see \cite{JT2011} for a
review of the momentum literature. Interestingly, \cite{SJ2022} recently
demonstrated that momentum in individual stock returns can be attributed to
momentum in other risk factors. This finding challenges the widely accepted
view in the literature that momentum is an independent risk factor.


One of the key findings in \cite{SJ2022} is that momentum in factor returns
effectively explains various forms of individual stock momentum, including
standard stock return momentum, industry momentum, industry-adjusted
momentum, intermediate momentum, and Sharpe ratio momentum. Specifically,
when regressing the monthly returns of individual stock momentum strategies
on the Fama-French five factors and factor momentum, the authors demonstrate
that none of the alphas in the asset pricing model are statistically
significant. Conversely, individual stock factors and the Fama-French five
factors fail to account for the abnormal returns of factor momentum.%
\footnote{%
For more details, see Table 5 and the corresponding discussions in \cite%
{SJ2022}.}

In this empirical application, we further shed light on the interaction
between individual stock momentum and factor momentum by estimating the $%
\mathrm{MSS}$ for various combinations of momentum strategies. We address
the following two research questions. First, do the individual stock
momentum factor and the factor momentum independently contribute to the
mean-variance efficiency when analyzed alongside other well-established
factors? Second, what role does factor momentum play in constructing the
mean-variance efficient frontier when it coexists with the individual stock
momentum factor? In contrast to \cite{SJ2022}, our analysis emphasizes the
relative importance of various momentum strategies in shaping the
mean-variance efficient frontier within the corresponding factor portfolios.
If a momentum strategy is more likely to be included in the $\mathrm{MSS}$,
it should be regarded as a critical component in mean-variance analysis.

\subsection{Data\label{sec:data}}

Our empirical application utilizes the same dataset of monthly factor
returns in the U.S. stock market as employed in \cite{SJ2022}. As noted in
the original study, the factor data are sourced from three primary
providers: Kenneth French's website, AQR, and Robert Stambaugh.\footnote{%
These factor returns are available for download at %
\url{http://mba.tuck.dartmouth.edu/pages/faculty/ken.french/data_library.html}%
, \url{https://www.aqr.com/insights/datasets}, and %
\url{https://finance.wharton.upenn.edu/~stambaug/}.} When not directly
available, factor returns are computed as the difference between the average
returns of the top three deciles and the bottom three deciles. The
construction of these decile portfolios strictly follows the methodology
outlined in the corresponding reference.

We study the $\mathrm{MSS}$ for five individual stock momentum factors and
two factor momentum strategies analyzed in \cite{SJ2022}. The five
individual stock momentum factors include the standard individual stock
momentum of \cite{JT1993}, the industry-adjusted momentum of \cite{CP1998},
the industry momentum of \cite{MG1999}, the intermediate momentum of \cite%
{NM2012}, and the Sharpe ratio momentum of \cite{RTSF2007}.\footnote{%
These individual stock momentum factors are constructed using the common
up-minus-down method with return-sorted portfolios. Specifically, the
standard individual stock momentum sorts stocks based on prior returns over
the past four quarters, the industry-adjusted momentum sorts stocks by prior
industry-adjusted returns, intermediate momentum sorts stocks by returns
from month $t-12$ to $t-7$, the Sharpe ratio momentum sorts stocks by
returns scaled by return volatility, and industry momentum sorts 20
industries based on their prior six-month returns.} The two factor momentum
strategies are the time-series momentum applied to 20 \textquotedblleft
off-the-shelf\textquotedblright \ individual factors and the momentum in the
first 10 principal component (PC) factors extracted from the 47 factors
studied in \cite{KNS2020}. For further details on the construction of these
factor momentum strategies, refer to \cite{SJ2022}. Panel A of Table \ref%
{tab:factors} presents references and summary statistics for these seven
momentum-related factors.

In addition, we consider a comprehensive set of well-established factors in
our analysis of the $\mathrm{MSS}$ of momentum strategies, including excess
market return, size, value, profitability, investment, accruals, betting
against beta, cash flow to price, earnings to price, liquidity, long-term
reversals, net share issues, quality minus junk, residual variance, and
short-term reversal. Panel B of Table~\ref{tab:factors} provides the
corresponding references and summary statistics for these factors.

Turning to the descriptive statistics reported in Table \ref{tab:factors},
we observe significant variations in the return performance of both momentum
strategies and other factors. For example, for the two factor momentum
strategies (FTM and PCM), the momentum in individual factors demonstrates a
substantially higher monthly return (0.33\% vs. 0.19\%) and volatility
(1.20\% vs. 0.64\%) compared to the momentum in the first 10 principal
component factors, whereas the average monthly returns of individual stock
factors are generally higher than those of factor momentum strategies,
albeit with significantly greater monthly volatility. Meanwhile, the
betting-against-beta factor generates the highest monthly return of 0.88\%
with a monthly volatility of 3.34\%.

\begin{table}[htbp]
\caption{Details on Stock Momentums, Factor Momentums, and other Factors}
\label{tab:factors}\centering {\scriptsize \ 
\setlength \extrarowheight{-2.0pt}
	\renewcommand{\arraystretch}{1.5} 
\begin{threeparttable}
		\begin{tabular}{lllccc}\toprule
			Factors  & Acronym	& Reference	&  Sample Period   &  Mean  & S.D.   \\  \hline
\multicolumn{6}{c}{\textbf{Panel A: Momentum Factors}} 		\\  
Momentum in individual factors  &  \text{FTM}	& \cite{SJ2022}  & 1964/07-2019/12  &  0.33\% &    1.20\%		\\  
Momentum in PC factors 1-10  & \text{PCM} & \cite{SJ2022}  &  1973/07-2019/12  & 0.19\%  &    0.64\%		\\  	
Standard stock momentum  &  \text{MOM}	& \cite{JT1993}  & 1964/07-2019/12  & 0.64\%  &   4.22\%	\\  
Industry-adjusted momentum  &  \text{IAM}	& \cite{CP1998} & 1964/07-2019/12  &  0.41\% &  2.64\%		\\  
Industry momentum  & \text{IDM}	& \cite{MG1999}  & 1964/07-2019/12  & 0.63\%  &   4.60\%		\\  
Intermediate momentum   &	\text{ITM} & \cite{NM2012}  & 1964/07-2019/12  & 0.48\%  &   3.02\%		\\  
Sharpe ratio momentum  &	 \text{SRM} & \cite{RTSF2007}  & 1964/07-2019/12  & 0.55\%  &  3.59\%		\\  
			\multicolumn{6}{c}{\textbf{Panel B: Other Factors}} 		\\  
			Excess market return  &	\text{MKT}  &  \cite{SR1964} &  1964/07-2019/12 & 0.53\%  &   4.42\%	\\  
			Size  &	\text{SMB}  &  \cite{Rolf1981} &  1964/07-2019/12 & 0.24\%  &   3.03\%	\\  
			Value & \text{HML}	&  \cite{BKR1985}  &  1964/07-2019/12  &  0.29\%  &   2.83\% 		\\  
			Profitability  &  \text{RMW}	& \cite{Robert2013}  &  1964/07-2019/12  &  0.27\%  &   2.17\%	\\  
			Investment   & \text{CMA}	& \cite{TWX2004}  & 1964/07-2019/12 &  0.27\%  &   2.00\% 		\\  
			Accruals  &	\text{ACC} & \cite{Sloan1996} & 1964/07-2019/12 &  0.22\%  &   1.91\%	\\ 
			Betting against beta  & \text{BAB}	&  \cite{FP2014}  & 1964/07-2019/12  & 0.83\%  &   3.27\%		\\  
			Cash flow to price   & \text{CFP}	&  \cite{BKR1985} & 1964/07-2019/12 & 0.27\%  &   2.51\%		\\  
			Earnings to price   & \text{ETP} &  \cite{Basu1983}  & 1964/07-2019/12 & 0.29\%  &   2.58\%		\\  
			Liquidity   & \text{LIP}	& \cite{PLS2003}  & 1968/01-2019/12 &  0.37\%  &    3.34\%		\\  
			Long-term reversals   &  \text{LTR}	& \cite{DT1985}  &  1964/07-2019/12  & 0.21\%  &   2.52\%		\\  
			Net share issues  &	\text{NSI}  & \cite{LR1995}  & 1964/07-2019/12  & 0.22\%  &   2.38\% 		\\  
			Quality minus junk   & \text{QMJ}	& \cite{AFP2019} &  1964/07-2019/12  &  0.39\% &   2.25\%		\\  
			Residual variance  &  \text{RVA}	& \cite{AHXZ2006}  & 1964/07-2019/12  & 0.12\%  &    5.02\%		\\  
			Short-term reversals  &	 \text{STR}  & \cite{Jegadeesh1990}  & 1964/07-2019/12  & 0.50\%  &   3.09\% 		\\ 			
			\bottomrule
		\end{tabular}
		\begin{tablenotes}[flushleft]
			\item \textit{Note:} This table provides the references, acronyms, sample periods, means, and standard deviations of the monthly returns for seven momentum factors and 15 U.S. factors examined in the empirical analysis.
		\end{tablenotes}
	\end{threeparttable}
}
\end{table}

\subsection{Empirical Results}

We first examine whether individual stock momentum and factor momentum
contribute to the mean-variance efficient frontiers formed by the well-known
Fama-French five factors. This assessment is accomplished by estimating the $%
\mathrm{MSS}$ for portfolios that combine one momentum factor with the five
factors. The findings, as presented in Table \ref{tab:MSS_momentum_1},
reveal consistent results across all factor combinations: the momentum
factor, whether individual stock momentum or factor momentum, is
consistently selected alongside the market, size, value, profitability, and
investment factors. These results provide robust evidence that momentum
factors significantly enhance the mean-variance efficient frontier
established by the Fama-French five factors, corroborating the existing
literature on stock return momentum; see \cite{JT1993, carhart1997,
RTSF2007, SJ2022}, among others.

\begin{table}[htbp]
\caption{MSS for a Single Momentum Factor and the Fama-French Five Factors}
\label{tab:MSS_momentum_1}
\centering {\footnotesize \  
\setlength \extrarowheight{-2.0pt}
		\renewcommand{\arraystretch}{2}  
\begin{threeparttable} 
			\begin{tabularx}{1\textwidth}{lYYYYYYY}\toprule
				Definition	  &  FTM	&  PCM  &  MOM  &  IAM & IDM &  ITM & SRM  \\  \hline
				Momentum factor & \Checkmark   & \Checkmark	&  \Checkmark  & \Checkmark  & \Checkmark  & \Checkmark   &  \Checkmark  \\ 
				Excess market return  &  \Checkmark  & \Checkmark & \Checkmark  & \Checkmark  & \Checkmark  &  \Checkmark  &  \Checkmark  \\ 
				Size  &  \Checkmark  & \Checkmark	& \Checkmark & \Checkmark & \Checkmark & \Checkmark & \Checkmark \\ 
				Value  &   &  &  &  &  &   &  \\ 
				Profitability & \Checkmark & \Checkmark	& \Checkmark & \Checkmark & \Checkmark & \Checkmark & \Checkmark \\ 
				Investment & \Checkmark & \Checkmark & \Checkmark & \Checkmark & \Checkmark & \Checkmark & \Checkmark \\ 		
				\bottomrule
			\end{tabularx}
			\begin{tablenotes}[flushleft]
				\item \textit{Note:} This table presents the estimated MSS for portfolios comprising a single momentum factor and the Fama-French five factors. The momentum factor in the row corresponds to the acronyms listed in the respective columns. Detailed descriptions of the factors, their acronyms, and sampling periods are provided in Table \ref{tab:factors}. A check mark (\Checkmark) denotes that the corresponding asset is included in the estimated MSS. All analyses are performed at a significance level of 0.05.
			\end{tablenotes}
	\end{threeparttable} }
\end{table}

\begin{table}[htbp]
\caption{MSS for Stock Momentum, Factor Momentum, and the Fama-French Five
Factors}
\label{tab:MSS_momentum_2}
\centering {\footnotesize \  
\setlength \extrarowheight{-2.0pt}
		\renewcommand{\arraystretch}{2}  
\begin{threeparttable} 
			\begin{tabularx}{1\textwidth}{lYYYYY}\toprule
				Definition  &     \text{MON}    &  \text{IAM}   &  \text{IDM}  &  \text{ITM}  &  \text{SRM}    \\ \hline
				\multicolumn{6}{c}{\textbf{Panel A}} 	\\ 		
				Stock momentum factor  & \Checkmark   & 	&  \Checkmark  &   &  \Checkmark     \\ 		
				Momentum in individual factors &  \Checkmark  & \Checkmark	& \Checkmark   & \Checkmark  &    \Checkmark   \\  
				Excess market return  & \Checkmark   & \Checkmark	& \Checkmark   & \Checkmark  &  \Checkmark    \\ 
				Size  & \Checkmark   & \Checkmark	& \Checkmark   & \Checkmark  &   \Checkmark    \\ 
				Value  & \Checkmark   &  &  \Checkmark &  &     \\ 
				Profitability &  \Checkmark  & \Checkmark	& \Checkmark   &  \Checkmark &   \Checkmark   \\ 
				Investment &  \Checkmark  & \Checkmark	&  \Checkmark  & \Checkmark &  \Checkmark    \\ 	\midrule			
				\multicolumn{6}{c}{\textbf{Panel B}} 	\\  
				Stock momentum factor  & \Checkmark   & \Checkmark	&  \Checkmark  &  \Checkmark &  \Checkmark     \\ 			
				Momentum in PC factors 1–10  & \Checkmark & \Checkmark &\Checkmark& \Checkmark  &  \Checkmark \\ 	 
				Excess market return  & \Checkmark & \Checkmark	&  \Checkmark  & \Checkmark  &  \Checkmark     \\ 
				Size  &  \Checkmark  & \Checkmark	&  \Checkmark  & \Checkmark  &       \\ 
				Value  &   & 	&    &   &     \\ 
				Profitability &  \Checkmark & \Checkmark	&  \Checkmark  & \Checkmark  & \Checkmark     \\ 
				Investment &  \Checkmark & \Checkmark	&  \Checkmark  & \Checkmark  &  \Checkmark     \\ 			
				\bottomrule
			\end{tabularx}
			\begin{tablenotes}[flushleft]
				\item \textit{Note:} This table presents the estimated MSS for portfolios comprising a single stock momentum factor, a single factor momentum strategy, and the Fama-French five factors. Panel A evaluates sets that include individual momentum factors, while Panel B examines sets incorporating the first 10 principal component factors. In both Panels A and B, the stock momentum factor in the row corresponds to the acronyms listed in the respective columns. Detailed descriptions of the factors, their acronyms, and sampling periods are provided in Table \ref{tab:factors}. A check mark (\Checkmark) indicates that the corresponding asset is included in the estimated MSS. All analyses are conducted at a significance level of 0.05.
			\end{tablenotes}
	\end{threeparttable} }
\end{table}

To distinguish the relative importance of individual stock momentum and
factor momentum in shaping the efficient frontier, we next estimate the $%
\mathrm{MSS}$ for portfolios comprising one stock momentum strategy, one
factor momentum strategy, and the Fama-French five factors. Table \ref%
{tab:MSS_momentum_2} reports the estimation results, highlighting three key
findings. First, factor momentum is included in the estimated $\mathrm{MSS}$
for all ten cases, underscoring its critical role in constructing efficient
frontiers. Second, although factor regressions suggest that individual stock
momentum is largely explained by factor momentum \citep{SJ2022}, it
nonetheless contributes to mean-variance efficient frontiers. Third, the
market, size, profitability, and investment factors remain relevant to
efficient frontiers when either momentum in individual factors or momentum
in the first ten principal component factors (PCM) is present, whereas the
contribution of the value factor is limited.

Our subsequent analysis evaluates the $\mathrm{MSS}$ for portfolios
consisting solely of momentum factors, delineating which factors are most
influential for the efficient frontier. Panel A of Table \ref%
{tab:MSS_momentum_3} details the estimated $\mathrm{MSS}$, consistently
including the factor momentum strategy. However, only the PCM is selected
when both factor momentum strategies are present. Additionally, various
individual stock momentum factors such as standard stock momentum (MOM),
industry-adjusted momentum (IAM), and Sharpe ratio momentum (SRM)
significantly contribute to the efficient frontier when considering all
related momentum factors. This finding contrasts with the findings of \cite%
{SJ2022}, which show that factor momentum fully accounts for stock return
momentum. Instead, our results underscore the distinctive contribution of
individual stock momentum in shaping mean-variance efficient frontiers.

\begin{figure}[H]
\caption{Graphical Illustration of the MSS Estimation Procedure}
\label{fig:diag_MSS_2}
\centering 	
\vspace{-0.5cm}  \includegraphics[width=0.95\textwidth]{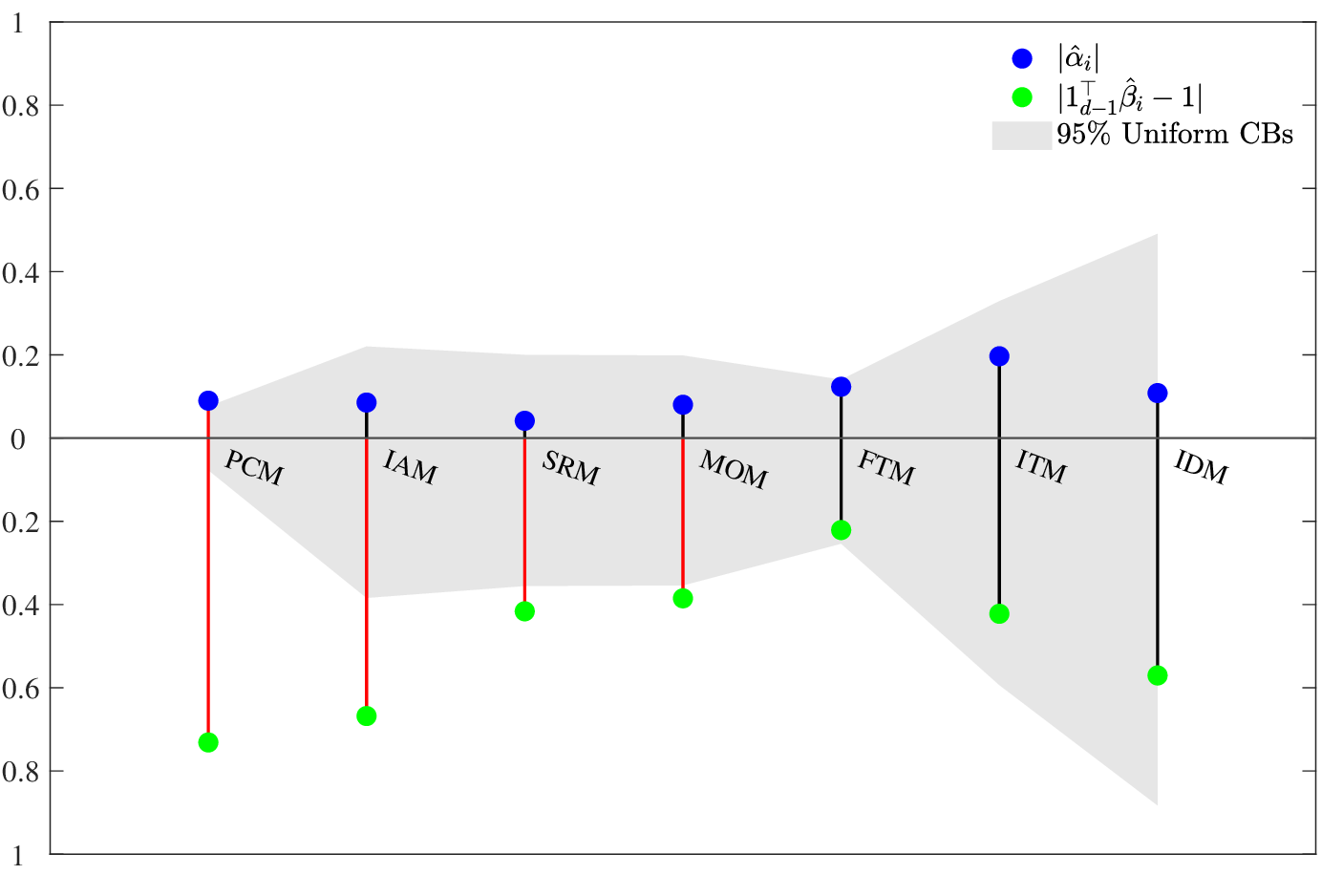} 
\newline
\begin{minipage}[t]{0.95 \textwidth}
		\small \textit{Note:} This figure illustrates the estimation procedure for the $\mathrm{MSS}$, as analyzed in the last column of Panel A in Table \ref{tab:MSS_momentum_3}. The assets are ordered from left to right based on their $M_{i,T}$ values, arranged in descending order (largest to smallest). The blue points above the $x$-axis represent the values of $|\hat{\alpha}_i|$, while the green points below the $x$-axis represent the values of $|1^\top_{d-1}\hat{\beta}_i - 1|$. The shaded regions above and below the $x$-axis indicate the 95\% uniform confidence bands for $|\hat{\alpha}_i|$ and $|1^\top_{d-1}\hat{\beta}_i - 1|$, respectively. An asset is included in the $\mathrm{MSS}$ if, and only if, $|\hat{\alpha}_i|$, $|1^\top_{d-1}\hat{\beta}_i - 1|$, or both exceed their respective 95\% uniform confidence bands. In such cases, the corresponding lines are highlighted in red.
	\end{minipage}
\end{figure}

Our estimation procedure provides more detailed insights into the relative
importance of assets within the full set compared to the information
presented in the table. To illustrate, Figure \ref{fig:diag_MSS_2} displays
plots of $|\hat{\alpha}_{i}|$ and $|1_{d-1}^{\top }\hat{\beta}_{i}-1|$ along
with their uniform confidence bands, for all momentum factors corresponding
to the third column of Panel A in Table \ref{tab:MSS_momentum_3}. The
uniform confidence bands for $|\hat{\alpha}_{i}|$ and $|1_{d-1}^{\top }\hat{%
\beta}_{i}-1|$ start at zero, with their upper bounds defined as $%
T^{-1/2}cv_{0.95,T}^{b}\hat{s}_{i,1}^{b}$ and $T^{-1/2}cv_{0.95,T}^{b}\hat{s}%
_{i,2}^{b}$, respectively. According to the estimation procedure, an asset
is included in the $\mathrm{MSS}$ if, and only if, $|\hat{\alpha}_{i}|$ or $%
|1_{d-1}^{\top }\hat{\beta}_{i}-1|$, or both exceed their respective
confidence bands.

From Figure \ref{fig:diag_MSS_2}, we observe that among the four momentum
factors included in the estimated MSS, PCM is the most significant one. This
is evident as both its $|\hat{\alpha}_{i}|$ and $|1_{d-1}^{\top }\hat{\beta}%
_{i}-1|$ are substantially different from zero. For the other three momentum
factors in the estimated MSS, their inclusion is driven by $|1_{d-1}^{\top }%
\hat{\beta}_{i}-1|$ being significantly different from zero, even though
their $|\hat{\alpha}_{i}|$ values remain within the uniform confidence
bands. Based on the statistic $M_{i,T}$, the ranking of momentum factors in
the estimated MSS is as follows: PCM, IAM, SRM, and MOM.

\begin{landscape}
	\begin{figure}[htbp!]
		\caption{Graphical Illustration of the MSS Estimation Procedure}
		\label{fig:diag_MSS}
		\centering 	
		\vspace{-0.5cm}
		\includegraphics[width=1.2\textwidth]{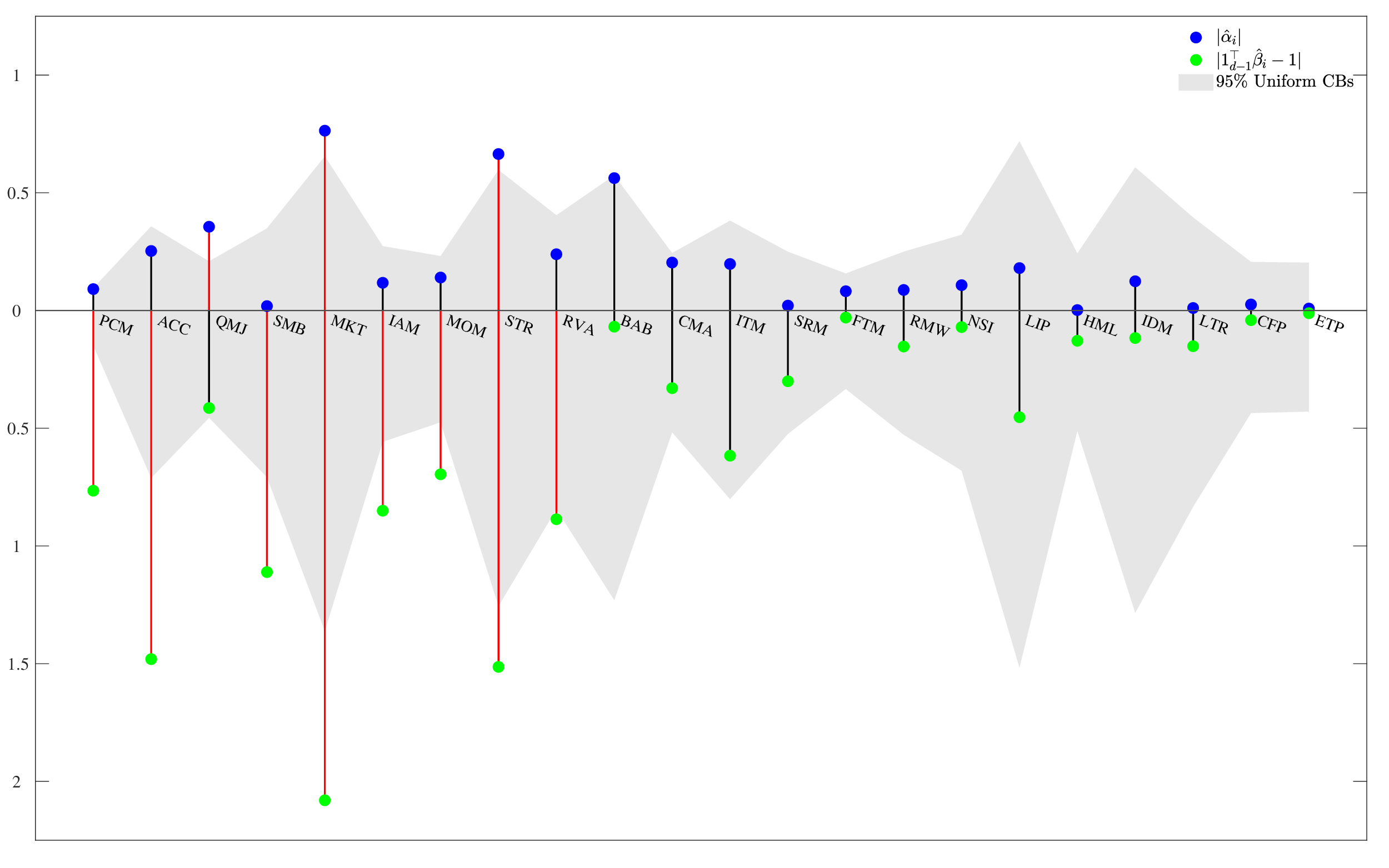} \newline
		\begin{minipage}[t]{1.2 \textwidth}
			\small \textit{Note:} This figure illustrates the estimation procedure for the MSS, as analyzed in the last column of Panel B in Table \ref{tab:MSS_momentum_3}. The assets are ordered from left to right based on their $M_{i,T}$ values, arranged in descending order (largest to smallest). The blue points above the $x $-axis represent the values of $|\hat{\alpha}_i| $, while the green points below the $%
			x $-axis depict the values of $|1^\top_{d-1}\hat{\beta}_i - 1| $. The shaded
			regions above and below the $x $-axis denote the 95\% uniform confidence
			bands for $|\hat{\alpha}_i| $ and $|1^\top_{d-1}\hat{\beta}_i - 1|$, respectively. An asset is included in the MSS if, and only if, $|\hat{\alpha}_i|$, $|1^\top_{d-1}\hat{\beta}_i - 1|$, or both exceed their respective 95\% uniform confidence bands. In such cases, the corresponding lines are highlighted in red.
		\end{minipage}
	\end{figure}
\end{landscape}

Finally, we analyze the $\mathrm{MSS}$ for a comprehensive pooled portfolio
comprising momentum factors, the Fama-French five factors, and ten
additional well-established factors, as detailed in Panel B of Table \ref%
{tab:factors}. The estimation results are provided in Panel B of Table \ref%
{tab:MSS_momentum_3}. To offer a graphical representation and further
evaluate the relative importance of the factors included in the estimated
MSS, we present plots of $|\hat{\alpha}_{i}|$ and $|1_{d-1}^{\top }\hat{\beta%
}_{i}-1|$ along with their uniform confidence bands, for each factor in the
full set in Figure \ref{fig:diag_MSS}, which corresponds to the results
shown in the third column of Panel B in Table \ref{tab:MSS_momentum_3}.

\begin{table}[]
\caption{MSS for Momentum Factors and Other Well-Established Factors}
\label{tab:MSS_momentum_3}
\centering {\footnotesize \ 
\setlength \extrarowheight{-2.0pt}
	\renewcommand{\arraystretch}{1.35}  
\begin{threeparttable}
		\begin{tabularx}{1\textwidth}{lYYY}\toprule
			Definition  &  \text{FTM}   & \text{PCM}   &   \text{FTM+PCM}    \\  \hline 
		 	\multicolumn{4}{c}{\textbf{Panel A}} 	\\ 	
			Momentum in individual factors  & \Checkmark  &  N.I.	&        \\ 
			Momentum in PC factors 1–10  & N.I.  & \Checkmark	& \Checkmark   \\ 
			Standard momentum  & \Checkmark  & \Checkmark	&  \Checkmark     \\ 
			Ind.-adjusted momentum  & \Checkmark  & \Checkmark	&  \Checkmark     \\ 
			Industry momentum   &   & 	&         \\ 
			Intermediate momentum  & \Checkmark  & 	&        \\ 
			Sharpe ratio momentum  &   & \Checkmark	&  \Checkmark     \\  \hline
			\multicolumn{4}{c}{\textbf{Panel B}} 	\\ 	
			Momentum in individual factors  & \Checkmark  & N.I.	&       \\ 
Momentum in PC factors 1–10  &  N.I. & \Checkmark	&    \Checkmark     \\  
Standard momentum  & \Checkmark  & \Checkmark	&    \Checkmark     \\ 
Ind.-adjusted momentum  & \Checkmark  & \Checkmark	&   \Checkmark      \\ 
Industry momentum   &   & 	&        \\ 
Intermediate momentum  & \Checkmark  & 	&        \\ 
Sharpe ratio momentum  &   & 	&      \\ 
			Excess market return   & \Checkmark  & \Checkmark	&    \Checkmark       \\ 
			Size   & \Checkmark  & \Checkmark	&   \Checkmark      \\ 
			Value   &  & 	&      \\ 
			Profitability  &   & 	&         \\ 
			Investment  & \Checkmark  & 	&         \\ 
			Accruals   & \Checkmark  & \Checkmark	&     \Checkmark     \\ 
			Betting against beta   &   & 	&           \\ 
			Cash flow to price    &  & 	&        \\ 
			Earnings to price   &   & 	&        \\ 
			Liquidity   &   & 	&      \\  
			Long-term reversals    &   & 	&      \\ 
			Net share issues   &   & 	&        \\ 
			Quality minus junk    & \Checkmark  & \Checkmark	&   \Checkmark    \\ 
			Residual variance   &   & \Checkmark	&   \Checkmark    \\ 
			Short-term reversals   & \Checkmark  & \Checkmark	&   \Checkmark     \\ 
			\bottomrule
		\end{tabularx}
		\begin{tablenotes}[flushleft]
			\item \textit{Note:} This table presents the estimated MSS for portfolios that include momentum factors and other factors. Panel A includes all momentum factors, while Panel B extends the analysis to portfolios comprising all momentum factors, the Fama-French five factors, and ten additional well-established factors. ``N.I.'' denotes that the corresponding factor is not included in the full set. Detailed descriptions of the factors, their acronyms, and sampling periods are provided in Table \ref{tab:factors}. A check mark (\Checkmark) indicates that the corresponding factor is included in the estimated MSS. All analyses are performed at a significance level of 0.05.
		\end{tablenotes}
	\end{threeparttable}
}
\end{table}

From Panel B of Table \ref{tab:MSS_momentum_3}, we observe that the size of
the estimated $\mathrm{MSS}$ is considerably smaller than the total number
of factors, highlighting the effectiveness of the proposed estimation
procedure in reducing the number of assets required to replicate the
mean-variance efficient frontier. Additionally, the estimated $\mathrm{MSS}$
consistently includes momentum factors across all three cases. Notably, PCM
subsumes FTM when both are present in the full set. Moreover, MOM and IAM
consistently contribute to the efficient frontier, whereas SRM, which
appears in the estimated $\mathrm{MSS}$ in several cases in Panel A, is
excluded in Panel B due to the presence of new factors in the full set.
Furthermore, the estimated $\mathrm{MSS}$ also consistently includes excess
market return (MKT), size (SMB), accruals (ACC), quality minus junk (QMJ),
and short-term reversal (STR), underscoring their significance in
mean-variance analysis.

From Figure \ref{fig:diag_MSS}, we observe that MKT and STR are included in
the estimated $\mathrm{MSS}$ because both their $|\hat{\alpha}i|$ values and
their $|1^\top{d-1}\hat{\beta}i - 1|$ values are significantly different
from zero. In contrast, PCM, MOM, IAM, SMB, ACC, and RVA are included solely
because their $|1^\top{d-1}\hat{\beta}_i - 1|$ values are significantly
different from zero, even though their $|\hat{\alpha}_i|$ values fall within
the uniform confidence bands. Additionally, QMJ is selected exclusively due
to its $|\hat{\alpha}i|$ value being significantly above zero. Based on the
statistic $M{i,T}$, the factors included in the estimated $\mathrm{MSS}$ are
ranked as follows: PCM, ACC, QMJ, SMB, MKT, IAM, MOM, STR, and RVA. This
ranking further highlights PCM's critical contribution to mean-variance
efficiency.

In summary, our analyses in this empirical application reaffirm the
significant role of factor momentum in asset pricing. Factor momentum is
consistently included in the $\mathrm{MSS}$ across all portfolios that
incorporate individual stock momentum, factor momentum strategies, and other
well-established factors. Notably, only the momentum in the first ten
principal components contributes to the efficient frontier when coexisting
with individual factor momentum. These findings align closely with those
reported in \cite{SJ2022}. Additionally, certain individual stock momentum
factors, such as the standard momentum factor and the industry-adjusted
momentum factor, also enhance the construction of the mean-variance
efficient frontier alongside factor momentum. Our method not only estimates
the $\mathrm{MSS}$, but also provides valuable insights into the sources of
each asset's contribution to mean-variance efficiency and their relative
importance within the full set of assets.

\setcounter{equation}{0} \renewcommand{\theequation}{\thesection.\arabic{equation}}


\section{Conclusion\label{sec:conclusion}}

This paper proposes a method for estimating the $\mathrm{MSS}$, the smallest
subset of assets capable of replicating the mean-variance efficient frontier
of the full asset set. We derive the identification conditions for the $%
\mathrm{MSS}$\ and propose an estimation and inference procedure based on
these conditions. Under some regularity assumptions, we theoretically
demonstrate that the proposed $\mathrm{MSS}$ estimator covers the true $%
\mathrm{MSS}$ wpa1, and converges to the true $\mathrm{MSS}$ with a
probability reaching any pre-specified confidence level. A comprehensive set
of Monte Carlo simulations shows that the procedure performs well in finite
samples when the sample size is sufficiently large.

The proposed estimation and inference procedure is applied to analyze the $%
\mathrm{MSS}$ for a collection of individual stock momentum and factor
momentum strategies, alongside other well-known factors. Our empirical study
highlights the significant role of factor momentum from the perspective of
mean-variance analysis, as it is consistently included in the estimated $%
\mathrm{MSS}$, when coexisting with individual stock momentum factors. This
analysis also provides new insights into the relative importance of various
individual momentum strategies and factors.

The asymptotic properties of our method are established under the assumption
that the size of the full asset set, denoted as $d$ in the paper, is fixed.
An interesting avenue for future research is to generalize the asymptotic
framework to allow $d$ to grow with the sample size. Additionally,
estimating the $\mathrm{MSS}$ in high-dimensional settings, where $d$ may
exceed the sample size, presents a compelling research direction. The
identification conditions established in this paper provide a guidance for
designing valid estimation procedures in high-dimensional scenarios. We
leave these extensions and related questions for future investigation.

{
	\small
	\bibliographystyle{econometrica}
	\bibliography{strong}
}


\appendix

\setcounter{equation}{0}
\setcounter{theorem}{0}
\setcounter{proposition}{0}
\setcounter{lemma}{0}
\renewcommand{\theequation}{A.\arabic{equation}}
\renewcommand*{\thesection}{A.\arabic{section}}
\renewcommand*{\thesubsection}{\thesection.\arabic{subsection}}
\renewcommand*{\thelemma}{A\arabic{lemma}}
\renewcommand*{\theproposition}{A\arabic{proposition}}
\renewcommand*{\theassumption}{\arabic{assumption}}

\newpage

\begin{center}
	\textbf{\LARGE{Appendix}}
\end{center}


\section{Proofs of the Main Results \label{AP: Main}}

This section presents the conditions required to establish the main results
of the paper, namely Lemma \ref{ID} and Theorem \ref{T2}, along with their
proofs. We begin by outlining the key conditions.

\begin{assumption}
\label{A1} (i) $\left \{ R_{t}\right \} _{t}$ is strictly stationary\ and
strong mixing with mixing coefficient $\alpha _{j}$ satisfying $\alpha
_{j}\leq Ca^{j}$ for some $a\in (0,1)$; (ii) $\mathrm{E}[\left \vert
R_{i,t}\right \vert ^{r}]\leq C$ for some constant $r>4$; (iii)\ the
eigenvalues of\ $\mathrm{Var}(R_{t})$ are bounded away from zero; (iv)\ the
dimension of $R_{t}$, denoted as $d$, is fixed.
\end{assumption}

Assumption \ref{A1} specifies conditions on the dimension, dependence
structure, moment bounds, and variance-covariance matrix of the vector
return process $\left \{ R_{t}\right \} _{t}$. Specifically, Assumptions \ref%
{A1}(i, ii, iv) are useful for establishing the consistency and the joint
asymptotic normality of the estimators $(\hat{\theta}_{i})_{i\leq d}$.
Meanwhile, Assumptions \ref{A1}(ii, iii) ensure that the variance of $R_{t}$
is finite and non-singular, thereby guaranteeing that the mean-variance
efficient frontier of $R_{t}$ is well-defined, not trivially reducible, and
that the sufficient conditions of Lemma \ref{ID} are satisfied.

\begin{assumption}
\label{A2}\ (i) the eigenvalues of $\Omega _{d,i}\equiv \lim_{T\rightarrow
\infty }\mathrm{Var}(T^{-1/2}\sum_{t\leq T}e_{i,t})\ $are bounded away from
zero;\ (ii) $\log (T)^{2}(\ell T^{-1/2+1/\tilde{r}}+\ell ^{-1})=o(1)$\ for
some constant $\tilde{r}\in (4,r)$.
\end{assumption}

Assumption \ref{A2} provides sufficient conditions to establish the validity
of the bootstrap procedure for the estimation and inference of the \textrm{%
MSS}. Assumption \ref{A2}(i) ensures that the asymptotic variance-covariance
matrix of the estimator $\hat{\theta}_{i}$ is nonsingular for any $i\leq d$.
Assumption \ref{A2}(ii) imposes restrictions on the bandwidth $\ell $ used
in the MBB procedure, specifically requiring that $\ell $ grows slower than\ 
$\log (T)^{-2}T^{1/2-1/\tilde{r}}$ but faster than\ $\log (T)^{2}$.

We now present the proofs of Lemma \ref{ID} and Theorem \ref{T2}. Throughout
this section, $C$ denotes a positive constant that is independent of $i$, $t$
and $T$,\ and may vary from line to line.

\bigskip

\noindent \textsc{Proof of Lemma\  \ref{ID}.}\ We begin by observing that $%
\Sigma ^{-1}1_{d}(1_{d}^{\top }\Sigma ^{-1}1_{d})^{-1}$ and $\Sigma ^{-1}\mu
(1_{d}^{\top }\Sigma ^{-1}\mu )^{-1}$ represent the global minimum variance
portfolio and the tangency portfolio (with zero risk-free rate),
respectively. The claim of the lemma is proved in two separate cases.

First, consider the case that $\mu \neq a1_{d}$ for any $a\in \mathbb{R}$.
In this case, the Markowitz algorithm (see, e.g., Theorem 3.1 in \cite%
{CM1995}) shows that the minimum variance portfolio $x_{e,\mu _{p}}$ with a
target mean return $\mu _{p}\ $has a unique solution:%
\begin{equation}
x_{e,\mu _{p}}=\omega _{\mu _{p}}\frac{\Sigma ^{-1}\mu }{1_{d}^{\top }\Sigma
^{-1}\mu }+(1-\omega _{\mu _{p}})\frac{\Sigma ^{-1}1_{d}}{1_{d}^{\top
}\Sigma ^{-1}1_{d}},  \label{P_IDL_1}
\end{equation}%
where 
\begin{equation*}
\omega _{\mu _{p}}\equiv \frac{(1_{d}^{\top }\Sigma ^{-1}\mu )1_{d}^{\top
}\Sigma ^{-1}(\mu _{p}1_{d}-\mu \mathbf{)}}{(\mu ^{\top }\Sigma ^{-1}\mu
)(1_{d}^{\top }\Sigma ^{-1}1_{d}\mathbf{)-}(1_{d}^{\top }\Sigma ^{-1}\mu 
\mathbf{)}^{2}}.
\end{equation*}%
The efficient portfolio frontier of $R$ is traced by $x_{e,\mu _{p}}$ as $%
\mu _{p}$ varies over $\mu _{p}\geq \mu ^{\top }\Sigma
^{-1}1_{d}(1_{d}^{\top }\Sigma ^{-1}1_{d})^{-1}$. Therefore, the assets with
non-zero weights in either $\Sigma ^{-1}1_{d}(1_{d}^{\top }\Sigma
^{-1}1_{d})^{-1}$ or $\Sigma ^{-1}\mu (1_{d}^{\top }\Sigma ^{-1}\mu )^{-1}$,
or both, constitute the $\mathrm{MSS}$, as they form the minimal set of
assets required to construct both these portfolios, which together span the
efficient frontier of $R$. This implies: 
\begin{equation}
\mathrm{MSS}=\mathrm{Supp}_{\Sigma ^{-1}\mu }\cup \mathrm{Supp}_{\Sigma
^{-1}1_{d}}.  \label{P_IDL_2}
\end{equation}%
Since $\Sigma ^{-1}1_{d}$ and $\Sigma ^{-1}\mu $ are both well-defined and
non-zero, the existence and uniqueness of the $\mathrm{MSS}$ follow from (%
\ref{P_IDL_2}). From Lemma \ref{OLS_Coff}, we have%
\begin{equation}
\mathrm{Supp}_{\Sigma ^{-1}\mu }=\mathrm{Supp}_{(\alpha _{i})_{i\leq d}}%
\text{ \  \ and \  \ }\mathrm{Supp}_{\Sigma ^{-1}1_{d}}=\mathrm{Supp}%
_{(1_{d-1}^{\top }\beta _{i}-1)_{i\leq d}},  \label{P_IDL_3}
\end{equation}%
which, along with (\ref{P_IDL_2}), establishes that the $\mathrm{MSS}$
satisfies (\ref{ID_2}).

In the case that $\mu =a1_{d}$ for some $a\in \mathbb{R}$, the efficient
portfolio frontier of $R$ shrinks to $\Sigma ^{-1}1_{d}(1_{d}^{\top }\Sigma
^{-1}1_{d})^{-1}$. In this case 
\begin{equation}
\mathrm{MSS}=\mathrm{Supp}_{\Sigma ^{-1}1_{d}}=\mathrm{Supp}_{\Sigma
^{-1}\mu }\cup \mathrm{Supp}_{\Sigma ^{-1}1_{d}},  \label{P_IDL_4}
\end{equation}%
where the second equality holds because $\mathrm{Supp}_{\Sigma ^{-1}\mu }$
is either empty (when $a=0$) or identical to $\mathrm{Supp}_{\Sigma
^{-1}1_{d}}$ (when $a\neq 0$). Since $\Sigma ^{-1}1_{d}$ is well-defined and
non-zero, (\ref{P_IDL_4}) confirms that the $\mathrm{MSS}$ exists and is
unique. Combining the results from (\ref{P_IDL_3}) and (\ref{P_IDL_4}), we
deduce that the $\mathrm{MSS}$ satisfies (\ref{ID_2}).\hfill $Q.E.D.$

\bigskip

\noindent \textsc{Proof of Theorem\  \ref{T2}.}\ Let $\theta _{i,1}^{0}\equiv
0$ and $\theta _{i,2}^{0}\equiv 1$. By the triangle inequality%
\begin{equation}
\min_{i\in \mathrm{MSS}}M_{i,T}\geq \min_{i\in \mathrm{MSS}}\max_{j=1,2}%
\frac{\left \vert T^{1/2}(\theta _{i,j}-\theta _{i,j}^{0})\right \vert }{%
\hat{s}_{i,j}}-\max_{i\in \mathrm{MSS}}\max_{j=1,2}\frac{\left \vert T^{1/2}(%
\hat{\theta}_{i,j}-\theta _{i,j})\right \vert }{\hat{s}_{i,j}}.
\label{P_T2_1}
\end{equation}%
From Lemma \ref{I_L3}, it follows that 
\begin{equation}
\max_{i\in \mathrm{MSS}}\max_{j=1,2}\frac{\left \vert T^{1/2}(\hat{\theta}%
_{i,j}-\theta _{i,j})\right \vert }{\hat{s}_{i,j}}=O_{p}(1).  \label{P_T2_2}
\end{equation}%
Let $Q\equiv \mathrm{E}[\tilde{R}_{t}\tilde{R}_{t}^{\top }]$ where $\tilde{R}%
_{t}\equiv (1,(R_{i,t})_{i\leq d}^{\top })^{\top }$. The matrix $Q$ can be
expressed as:%
\begin{equation*}
Q=\left( 
\begin{array}{cc}
1 & \mathbf{0}_{1\times d} \\ 
\mathrm{E}[R_{t}] & I_{d}%
\end{array}%
\right) \left( 
\begin{array}{cc}
1 & \mathbf{0}_{1\times d} \\ 
\mathbf{0}_{d\times 1} & \mathrm{Var}(R_{t})%
\end{array}%
\right) \left( 
\begin{array}{cc}
1 & \mathrm{E}[R_{t}^{\top }] \\ 
\mathbf{0}_{d\times 1} & I_{d}%
\end{array}%
\right) .
\end{equation*}%
Therefore, by Assumptions\  \ref{A1}(ii, iii), we have%
\begin{equation}
C^{-1}\leq \lambda _{\min }(Q)\leq \lambda _{\max }(Q)\leq C,  \label{P_T2_3}
\end{equation}%
where $\lambda _{\min }(Q)$ and $\lambda _{\max }(Q)$ denote the smallest
and largest eigenvalues of $Q$, respectively. From (\ref{P_T2_3}), it
further follows that 
\begin{equation}
\min_{i\leq d}\sigma _{\varepsilon _{i}}^{2}\geq \lambda _{\min }(Q)>C.
\label{P_T2_4}
\end{equation}%
Combining the results in (\ref{P_T2_3}) and (\ref{P_T2_4}) with Lemma \ref%
{I_L1b}, we have%
\begin{eqnarray}
\min_{i\in \mathrm{MSS}}\max_{j=1,2}\frac{\left \vert T^{1/2}(\theta
_{i,j}-\theta _{i,j}^{0})\right \vert }{\hat{s}_{i,j}} &\geq
&(C^{-1}-o_{p}(1))\min_{i\in \mathrm{MSS}}\max_{j=1,2}\frac{\left \vert
T^{1/2}(\theta _{i,j}-\theta _{i,j}^{0})\right \vert }{\left \Vert
A_{j,\cdot }\right \Vert }  \notag \\
&\geq &(C^{-1}-o_{p}(1))T^{1/2}\min_{i\in \mathrm{MSS}}\left( \alpha
_{i}^{2}+\frac{(1_{d-1}^{\top }\beta _{i}-1)^{2}}{d-1}\right) ^{1/2},
\label{P_T2_5}
\end{eqnarray}%
which, along with (\ref{T2_1}), (\ref{P_T2_1}) and (\ref{P_T2_2}), implies
that 
\begin{equation}
\min_{i\in \mathrm{MSS}}M_{i,T}\geq (C^{-1}-o_{p}(1))T^{1/2}\min_{i\in 
\mathrm{MSS}}\left( \alpha _{i}^{2}+\frac{(1_{d-1}^{\top }\beta _{i}-1)^{2}}{%
d-1}\right) ^{1/2}.  \label{P_T2_6}
\end{equation}%
Let $cv_{1-p,T}^{b}\equiv cv_{1-p,T}^{b}(\mathcal{M}\backslash \widehat{%
\mathrm{MSS}}_{p})$. Since $cv_{1-p,T}^{b}(\mathcal{M})=O_{p}(1)$ by Lemma %
\ref{B_L7a} and $0\leq cv_{1-p,T}^{b}\leq cv_{1-p,T}^{b}(\mathcal{M})$, it
follows from\ (\ref{P_T2_6}) that%
\begin{eqnarray}
\min_{i\in \mathrm{MSS}}M_{i,T}-cv_{1-p,T}^{b} &\geq
&(C^{-1}-o_{p}(1))T^{1/2}\min_{i\in \mathrm{MSS}}\left( \alpha _{i}^{2}+%
\frac{(1_{d-1}^{\top }\beta _{i}-1)^{2}}{d-1}\right) ^{1/2}-O_{p}(1)  \notag
\\
&\geq &(C^{-1}-o_{p}(1))T^{1/2}\min_{i\in \mathrm{MSS}}\left( \alpha
_{i}^{2}+\frac{(1_{d-1}^{\top }\beta _{i}-1)^{2}}{d-1}\right) ^{1/2},
\label{P_T2_7}
\end{eqnarray}%
where the second inequality is by (\ref{T2_1}). From (\ref{P_T2_7}), we
obtain: 
\begin{equation}
\mathrm{P}\left( \mathrm{MSS}\subseteq \widehat{\mathrm{MSS}}_{p}\right) =%
\mathrm{P}\left( \min_{i\in \mathrm{MSS}}M_{i,T}>cv_{1-p,T}^{b}\right) \geq 
\mathrm{P}\left( C^{-1}-o_{p}(1)>0\right) \geq 1-o(1),  \label{P_T2_8}
\end{equation}%
which shows the first claim of the theorem.\footnote{%
Since the $\mathrm{MSS}$ is nonempty, (\ref{P_T2_8}) implies that $\widehat{%
\mathrm{MSS}}_{p}$ is nonempty wpa1. Consequently, the finite sample
adjustment described in the paragraph following the algorithm in Subsection %
\ref{subsec:est} is asymptotically negligible.}

To establish the second claim of the theorem, we first observe that since\ $%
\mathcal{S}\subseteq \mathcal{S}^{\prime }$ implies $\max_{i\in \mathcal{S}%
}M_{i,T}^{b}\leq \max_{i\in \mathcal{S}^{\prime }}M_{i,T}^{b}$, it follows
that 
\begin{equation}
cv_{1-p,T}^{b}(\mathcal{S})\leq cv_{1-p,T}^{b}(\mathcal{S}^{\prime }).
\label{P_T1_0}
\end{equation}%
From this and (\ref{P_T2_8}), we have $cv_{1-p,T}^{b}\leq cv_{1-p,T}^{b}(%
\mathrm{MSS}^{c})$ wpa1, where $\mathrm{MSS}^{c}\equiv \mathcal{M}\backslash 
\mathrm{MSS}$. Thus 
\begin{equation}
\mathrm{P}\left( \max_{i\in \mathrm{MSS}^{c}}M_{i,T}\leq
cv_{1-p,T}^{b}\right) \leq \mathrm{P}\left( \max_{i\in \mathrm{MSS}%
^{c}}M_{i,T}\leq cv_{1-p,T}^{b}(\mathrm{MSS}^{c})\right) +o(1).
\label{P_T2_9}
\end{equation}%
By Lemma \ref{B_L7}, there exists a positive sequence $\delta _{T}=o(1)$
such that%
\begin{equation}
\mathrm{P}\left( \max_{i\in \mathrm{MSS}^{c}}M_{i,T}\leq cv_{1-p,T}^{b}(%
\mathrm{MSS}^{c})\right) \leq \mathrm{P}\left( \max_{i\in \mathrm{MSS}%
^{c}}M_{i,T}\leq cv_{1-p+\delta _{T}}^{u}(\mathrm{MSS}^{c})\right) +o(1).
\label{P_T2_10}
\end{equation}%
Therefore for any $\varepsilon \in (0,1-p)$, we have%
\begin{equation}
\mathrm{P}\left( \max_{i\in \mathrm{MSS}^{c}}M_{i,T}\leq cv_{1-p,T}^{b}(%
\mathrm{MSS}^{c})\right) \leq \mathrm{P}\left( \max_{i\in \mathrm{MSS}%
^{c}}M_{i,T}\leq cv_{1-p+\varepsilon }^{u}(\mathrm{MSS}^{c})\right) +o(1),
\label{P_T2_11}
\end{equation}%
for sufficiently large $T$. By Lemma \ref{I_L3}, it follows that 
\begin{equation}
\lim_{T\rightarrow \infty }\mathrm{P}\left( \max_{i\in \mathrm{MSS}%
^{c}}M_{i,T}\leq cv_{1-p+\varepsilon }^{u}(\mathrm{MSS}^{c})\right) =\mathrm{%
P}\left( \max_{i\in \mathrm{MSS}^{c}}\tilde{M}_{i}\leq cv_{1-p+\varepsilon
}^{u}(\mathrm{MSS}^{c})\right) .  \label{P_T2_12}
\end{equation}%
This, along with (\ref{P_T2_9}), (\ref{P_T2_11}) and the definition of $%
cv_{1-p-\varepsilon }^{u}(\mathrm{MSS}^{c})$ shows that 
\begin{equation}
\mathrm{P}\left( \max_{i\in \mathrm{MSS}^{c}}M_{i,T}\leq
cv_{1-p,T}^{b}\right) \leq 1-p+\varepsilon +o(1),  \label{P_T2_13}
\end{equation}%
from which, we derive an upper bound for the probability $\mathrm{MSS}=%
\widehat{\mathrm{MSS}}_{p}$: 
\begin{eqnarray}
\underset{T\rightarrow \infty }{\lim \sup }\mathrm{P}\left( \mathrm{MSS}=%
\widehat{\mathrm{MSS}}_{p}\right) &=&\underset{T\rightarrow \infty }{\lim
\sup }\mathrm{P}\left( \max_{i\in \mathrm{MSS}^{c}}M_{i,T}\leq
cv_{1-p,T}^{b}\bigcap \min_{i\in \mathrm{MSS}}M_{i,T}>cv_{1-p,T}^{b}\right) 
\notag \\
&\leq &\underset{T\rightarrow \infty }{\lim \sup }\mathrm{P}\left(
\max_{i\in \mathrm{MSS}^{c}}M_{i,T}\leq cv_{1-p,T}^{b}\right) \leq
1-p+\varepsilon .  \label{P_T2_14}
\end{eqnarray}%
On the other hand, we can obtain a lower bound for the probability $\mathrm{%
MSS}=\widehat{\mathrm{MSS}}_{p}$ as follows: 
\begin{eqnarray}
\mathrm{P}\left( \mathrm{MSS}=\widehat{\mathrm{MSS}}_{p}\right) &=&\mathrm{P}%
\left( \mathrm{MSS}\subseteq \widehat{\mathrm{MSS}}_{p}\bigcap \widehat{%
\mathrm{MSS}}_{p}\subseteq \mathrm{MSS}\right)  \notag \\
&\geq &\mathrm{P}\left( \mathrm{MSS}\subseteq \widehat{\mathrm{MSS}}%
_{p}\right) -\mathrm{P}\left( \widehat{\mathrm{MSS}}_{p}\nsubseteq \mathrm{%
MSS}\right) .  \label{P_T2_15}
\end{eqnarray}%
In the proof of Lemma \ref{T1}, we have shown that%
\begin{equation*}
\underset{T\rightarrow \infty }{\lim \sup }\mathrm{P}\left( \widehat{\mathrm{%
MSS}}_{p}\nsubseteq \mathrm{MSS}\right) \leq p+\varepsilon .
\end{equation*}%
This, together with (\ref{P_T2_8}) and (\ref{P_T2_15}), implies that 
\begin{equation}
\underset{T\rightarrow \infty }{\lim \inf }\mathrm{P}\left( \mathrm{MSS}=%
\widehat{\mathrm{MSS}}_{p}\right) \geq 1-p-\varepsilon .  \label{P_T2_16}
\end{equation}%
Since $\varepsilon $ is arbitrary, the second claim of the theorem follows
from (\ref{P_T2_14}) and (\ref{P_T2_16}).\hfill $Q.E.D.$

\section{Auxiliary Lemmas \label{AP: AU_L}}

This section presents some auxiliary results that are utilized in proving
Lemma \ref{ID} and Theorem \ref{T2} in the previous section. For any $%
i=1,\ldots ,d$, let $\mu _{i}\equiv $ $\mathrm{E}[R_{i}]$, $\mu _{-i}\equiv $
$\mathrm{E}[R_{-i}]$, $\sigma _{i}^{2}\equiv \mathrm{Var}(R_{i})$ and $%
\Sigma _{-i}\equiv \mathrm{Var}(R_{-i})$. Define $\tilde{\sigma}%
_{i}^{2}\equiv \sigma _{i}^{2}-\Gamma _{i,-i}\Sigma _{-i}^{-1}\Gamma
_{i,-i}^{\top }$, where\ $\Gamma _{i,-i}\equiv \mathrm{Cov}(R_{i},R_{-i})$.

\begin{lemma}
\label{OLS_Coff}\ Suppose that $\Sigma $ is finite and non-singular.\ For
any $i=1,\ldots ,d$, the regression coefficients in (\ref{ID_1}) satisfy $%
\alpha _{i}=\tilde{\sigma}_{i}^{2}\ell _{d,i}^{\top }\Sigma ^{-1}\mu $ and $%
1_{d-1}^{\top }\beta _{i}-1=\tilde{\sigma}_{i}^{2}\ell _{d,i}^{\top }\Sigma
^{-1}1_{d}$, where $\tilde{\sigma}_{i}^{2}>0$.
\end{lemma}

\begin{lemma}
\label{I_L1b} Under Assumption \ref{A1}, we have $\max_{i\leq d}\max_{j=1,2}|%
\hat{s}_{i,j}^{2}/s_{i,j}^{2}-1|=O_{p}(T^{-1/2})$.
\end{lemma}

\begin{lemma}
\label{I_L3}\ Under Assumption \ref{A1},\ we have for any nonempty subset $%
\mathcal{S}$ of $\mathcal{M}$:%
\begin{equation*}
\max_{i\in \mathcal{S}}\max_{j=1,2}\frac{\left \vert T^{1/2}(\hat{\theta}%
_{i,j}-\theta _{i,j})\right \vert }{\hat{s}_{i,j}}\rightarrow _{d}\max_{i\in 
\mathcal{S}}\tilde{M}_{i},
\end{equation*}%
where $\tilde{M}_{i}\equiv \left \vert (\ell _{d,i}^{\top }\otimes \mathrm{%
diag}(s_{i,1}^{-1},s_{i,2}^{-1}))\Omega _{d}^{1/2}\mathcal{N}_{d}\right \vert
_{\infty }$\ and\ $\Omega _{d}=\lim_{T\rightarrow \infty }\Omega _{d,T}$.
\end{lemma}

For the $t$th observation $R_{t}^{b}$ in the bootstrap sample\ $%
\{R_{t}^{b}\}_{t\leq T_{B}}$, let $R_{-i,t}^{b}$ denotes its subvector
excluding $R_{i,t}^{b}$. Define $\tilde{R}_{-i,t}^{b}\equiv
(1,(R_{-i,t}^{b})^{\top })^{\top }$ and $\varepsilon _{i,t}^{b}\equiv
R_{i,t}^{b}-\theta _{i}^{\top }\tilde{R}_{-i,t}^{b}$. For any proper subset\ 
$\mathcal{S}\ $of $\mathcal{M}$, including the empty set, let\ $cv_{1-p}^{u}(%
\mathcal{S})$\ represent the\ ($1-p$)th quantile of $\max_{i\notin \mathcal{S%
}}\tilde{M}_{i}$,\ and\ $cv_{1-p,T}^{b}(\mathcal{S})$ represent the ($1-p$%
)th quantile of $\max_{i\notin \mathcal{S}}M_{i,T}^{b}$. Moreover, we
define: 
\begin{equation*}
\tilde{M}_{i,T}^{b}\equiv \max_{j=1,2}\frac{\left \vert
T_{B}^{-1/2}\sum_{t\leq T_{B}}e_{i,j,t}^{b}\right \vert }{s_{i,j}},
\end{equation*}%
with $e_{i,j,t}^{b}\equiv A_{j,\cdot }Q_{-i}^{-1}(\tilde{R}%
_{-i,t}^{b}\varepsilon _{i,t}^{b}-\mathrm{E}^{\ast }[\tilde{R}%
_{-i,t}^{b}\varepsilon _{i,t}^{b}])$ and $\mathrm{E}^{\ast }[\cdot ]$
denoting the expectation taken under the bootstrap distribution given the
data.

\begin{lemma}
\label{B_L7a} Under\ Assumptions \ref{A1} and \ref{A2}, we have $%
cv_{1-p,T}^{b}(\mathcal{M})=O_{p}(1)$ for any $p\in (0,1)$.
\end{lemma}

\begin{lemma}
\label{B_L7}\ Under\ Assumptions \ref{A1} and \ref{A2}, there exists a
positive sequence $\delta _{T}=o(1)$ such that for any $p\in (0,1)$,%
\begin{equation*}
cv_{1-p-\delta _{T}}^{u}(\mathcal{S})\leq cv_{1-p,T}^{b}(\mathcal{S})\leq
cv_{1-p+\delta _{T}}^{u}(\mathcal{S})\text{, \ wpa1.}
\end{equation*}
\end{lemma}

\begin{lemma}
\label{T1}\ Under Assumptions\  \ref{A1} and \ref{A2},\ we have\ $\lim
\inf_{T\rightarrow \infty }\mathrm{P}(\widehat{\mathrm{MSS}}_{p}\subseteq 
\mathrm{MSS})\geq 1-p$.
\end{lemma}

\end{document}